\documentclass[aps,prb,twocolumn,showpacs]{revtex4}
\usepackage{graphicx}
\usepackage{amssymb}%
\usepackage{epsfig}
\usepackage{amsmath}
\usepackage{ulem}
\usepackage{color}
\newcommand{\DM}{Dzyaloshinskii-Moriya }
\begin{document}
\title{A Schwinger-boson approach to the kagome with \DM interactions:\\phase diagram and dynamical structure factors} \author{L. Messio,$^1$ O.~C\'epas,$^{2}$ and C.~Lhuillier$^1$} \affiliation{$1.$
  Laboratoire de physique th\'eorique de la mati\`ere condens\'ee, UMR7600
  CNRS, Universit\'e Pierre-et-Marie-Curie, Paris 6, 75252 Paris cedex 05,
  France. \\ $2.$ Institut N\'eel, CNRS et Universit\'e Joseph Fourier, BP 166,
  F-38042 Grenoble Cedex 9, France. }
\date{\today}

\begin{abstract}
We have obtained the zero-temperature phase diagram of the kagom\'e
antiferromagnet with \DM interactions in Schwinger-boson mean-field
theory. We find quantum phase transitions (first or second order)
between different topological spin liquids and N\'eel ordered phases
(either the $\sqrt{3} \times \sqrt{3}$ state or the so-called $Q=0$
state). In the regime of small Schwinger-boson density, the results
bear some resemblances with exact diagonalization results and we
briefly discuss some issues of the mean-field treatment.  We calculate
the equal-time structure factor (and its angular average to allow for
a direct comparison with experiments on powder samples), which extends earlier work on the
classical kagom\'e to the quantum regime. We also discuss the dynamical
structure factors of the topological spin liquid and the N\'eel
ordered phase.
\end{abstract}

\maketitle

\section{Introduction}

The \DM interactions\cite{DM} are inevitably present in $S=1/2$ magnetic oxides
when the magnetic bonds lack inversion centers, which is the case of
the kagome lattice.  Although small in strength (it originates in
the spin-orbit coupling), such a correction may favour other phases
than the ones usually predicted by using the standard Heisenberg
model. By breaking explicitly the spin-rotation symmetry of the
system, the \DM forces tend to reduce the spin fluctuations and may
therefore be detrimental to the long-searched spin-liquid phases.
However, would the Heisenberg phase be gapped, such as in valence
bond crystals (generalized spin-Peierls states) or in topological spin
liquids,\cite{ReadSachdev91,Wen91,MisguichLhuillier2005} then it would be robust against perturbations
 typically smaller than the gap.  An example is given by the Shastry-Sutherland
compound SrCu$_2$(BO$_3$)$_2$ which remains in the singlet phase in
the presence of \DM couplings.\cite{Cepas0} In the kagome antiferromagnet (with pure nearest neighbor Heisenberg
interaction),  the very existence of a gap remains an open
question,\cite{Sindzingre} algebraic spin liquid and gapless vortex
phases have been proposed as alternatives in the recent
years.\cite{Hastings,Ran,Ryu} Current numerical
estimates of the
gap provide a small upper bound $\sim 0.05 J$.~\cite{Lecheminant,Waldtmann,Jiang,Sindzingre,Singhgap}  In any case the gap (if it exists)
could be smaller than the \DM coupling (especially in copper oxides
where it is typically $\sim 0.1J$) and the latter is therefore
particularly relevant. Experiments on the kagome compound
ZnCu$_3$(OH)$_6$Cl$_3$\cite{Mendels,Helton,deVries,deVries2} have
found no spin gap (despite a temperature much lower than the upper estimation of the gap), but the chemical
disorder\cite{Misguich,Imai,Olariu} and, precisely, the existence of
\DM interactions\cite{Zorko,Rigol} make the spin gap issue not yet clear.

In fact, the \DM interactions were argued to induce long-range $Q=0$,
120$^o$ N\'eel order in the kagome antiferromagnet: the algebraic spin
liquid theory predicts the instability at a critical strength
$D_c=0$,\cite{Hermele} while there is a finite quantum critical point
at $D_c \sim 0.1 J$ in exact diagonalization results on samples of
size up to $N=36$.\cite{Cepas} Since it is clear that there is no
N\'eel order at $D$ strictly zero \cite{Leung,Lecheminant} and there
is N\'eel order for $D$ large enough,\cite{Cepas} it is tempting to
tackle the problem using the Schwinger boson representation for the
spin operators.\cite{Auerbach} Indeed this approach allows in
principle to capture both topological spin liquid and N\'eel ordered
phases~\cite{Sachdev} and offers a first framework to describe this
quantum phase transition.  The caveat is that the actual
Schwinger-boson mean-field solution for the $S=1/2$, $D=0$ kagome
antiferromagnet is already long-ranged ordered, and it is only at
smaller values of $S$ (which in this approach is a continuous
parameter) that a disordered spin-liquid phase is stabilized. This
result may however be an artefact of the mean field approach, and it
is possible that fluctuations not taken into account at this level do
stabilize the disordered phase for the physical spin-1/2 system. It is
therefore interesting to see what phases the Schwinger-boson
mean-field theory predicts for the kagome antiferromagnet perturbed by
\DM interactions.

  In section ~\ref{Section:Model} we present the model and the method.
  In section~\ref{Section:phasediagram} we discuss the phase diagram
  together with general ground-state properties. In
  section~\ref{Section:spinon}, we illustrate the evolution
  of observables across the quantum phase transition
    from topological spin liquid to long-ranged N\'eel order: the
    spinon spectrum, the gap, the order-parameter and the condensed
    fraction of bosons. We calculate the equal-time structure factor,
    its powder average and briefly compare both to classical
    calculation and experimental results. In
  section~\ref{Section:dynamical} we present the dynamical spin
  structure factor and its behavior in the two phases.
  We also describe how these behaviors emerge from the
  properties of the spinon spectrum.

\section{Model}
\label{Section:Model}
\begin{figure}[h]
\centerline{
 \psfig{file=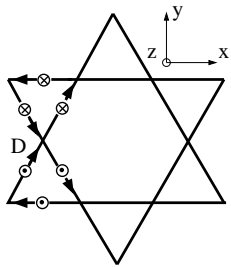,width=3cm,angle=0}
}
\caption{The \DM field in the model (spin rotated frame -see text). }
\label{figmodel}
\end{figure}

We have considered additional \DM interactions (DM) to  the standard Heisenberg model on the kagome lattice
\begin{equation}
H = \sum_{<i,j>} \left[ J \textbf{\mbox{S}}_i . \textbf{\mbox{S}}_j  + \textbf{\mbox{D}}
_{ij}. ( \textbf{\mbox{S}}_i \times \textbf{\mbox{S}}_j) \right]
\label{ham}
\end{equation}
where $<i,j>$ stands for nearest neighbours (each bond is counted
once) and $\mathbf{S}_i$ is a quantum spin operator on site $i$, the
\DM field $\textbf{\mbox{D}} _{ij}$ is taken to be along the $z$
 axis and is staggered from
up to down triangles (See Fig.~\ref{figmodel}). For
  this, we work in a
rotated frame which allows to eliminate
  the other components (the $\textbf{\mbox{S}}_i$ are to be viewed as
rotated operators).\cite{Cepas} In this case, the \DM field favors a
vector chirality along $z$ and the model (\ref{ham}) has a global
  $U(1)$ symmetry which can be spontaneously broken or not.

\begin{figure}[h]
\centerline{
 \psfig{file=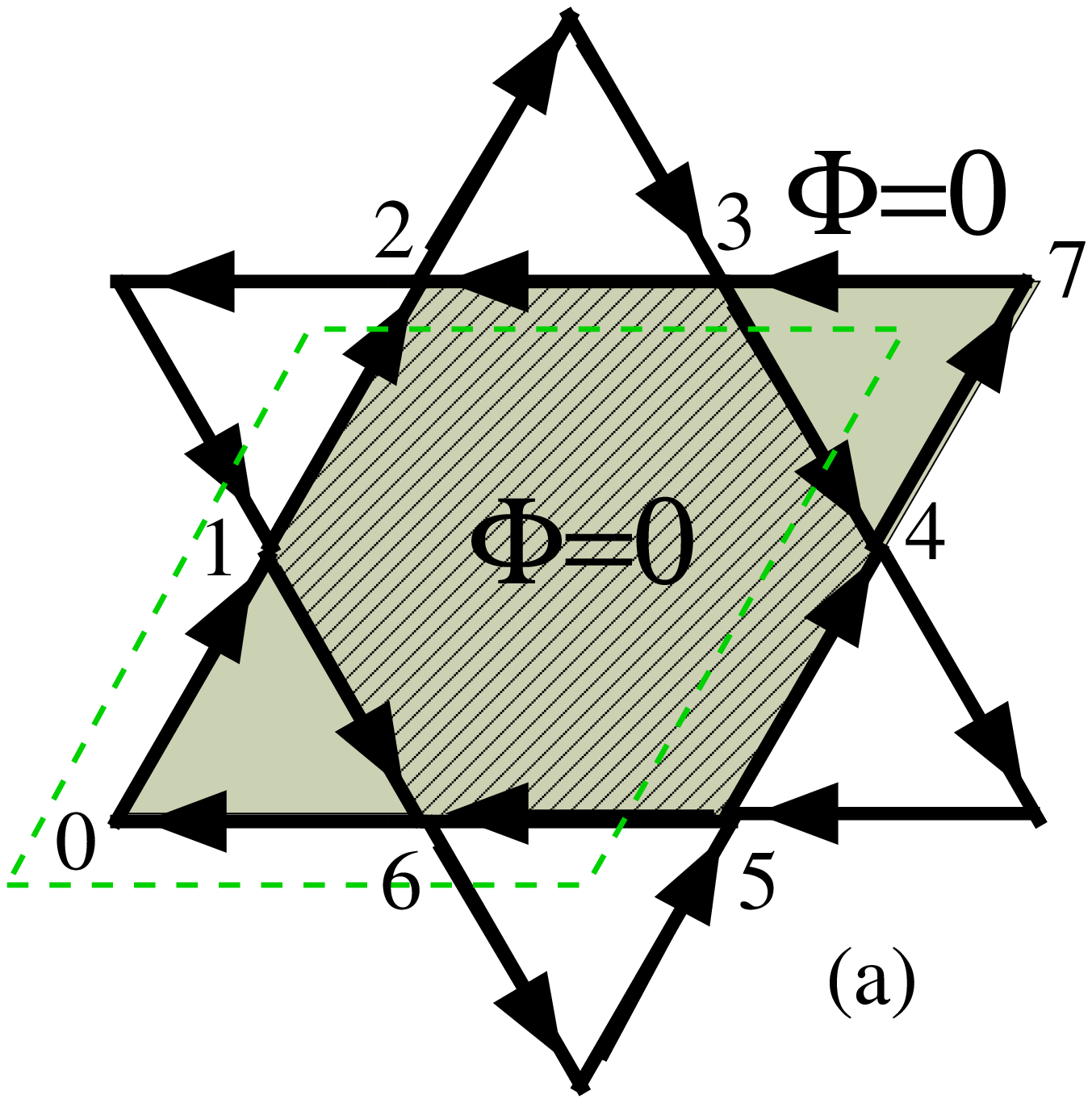,width=3cm,angle=0} \hspace{1cm}
 \psfig{file=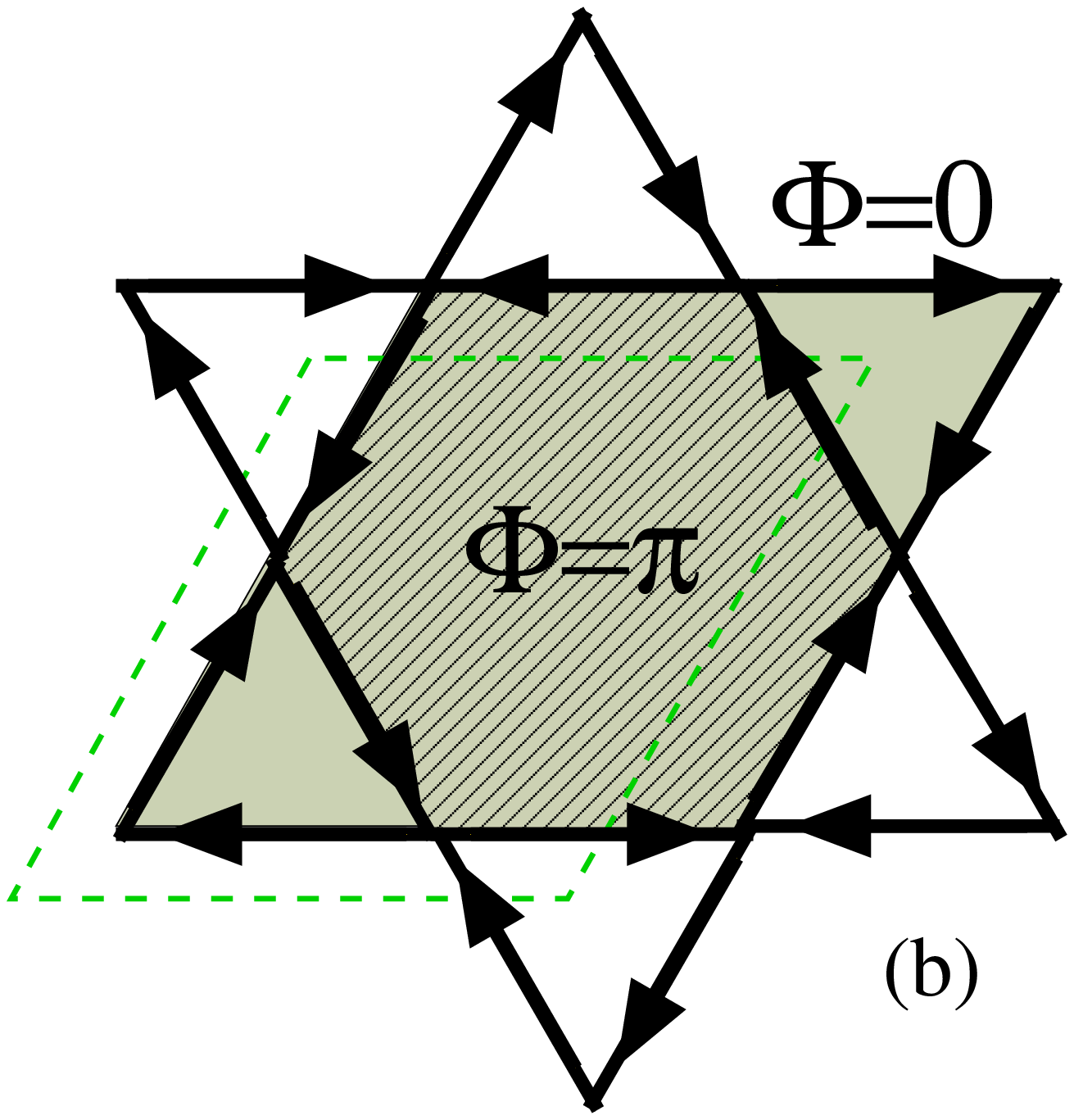,width=3cm,angle=0}
}
\vspace{.5cm}
\centerline{
 \psfig{file=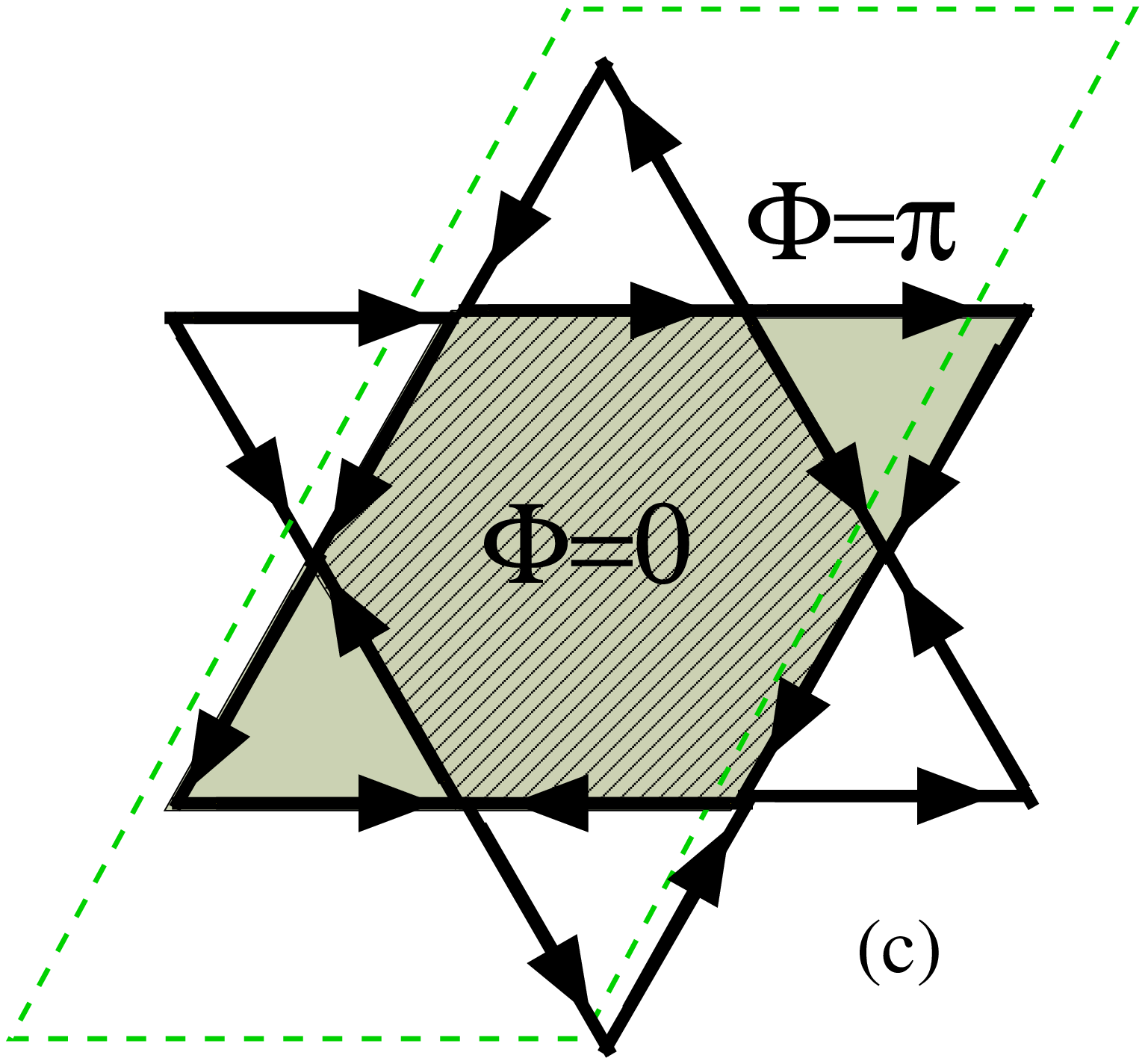,width=3cm,angle=0} \hspace{1cm}
 \psfig{file=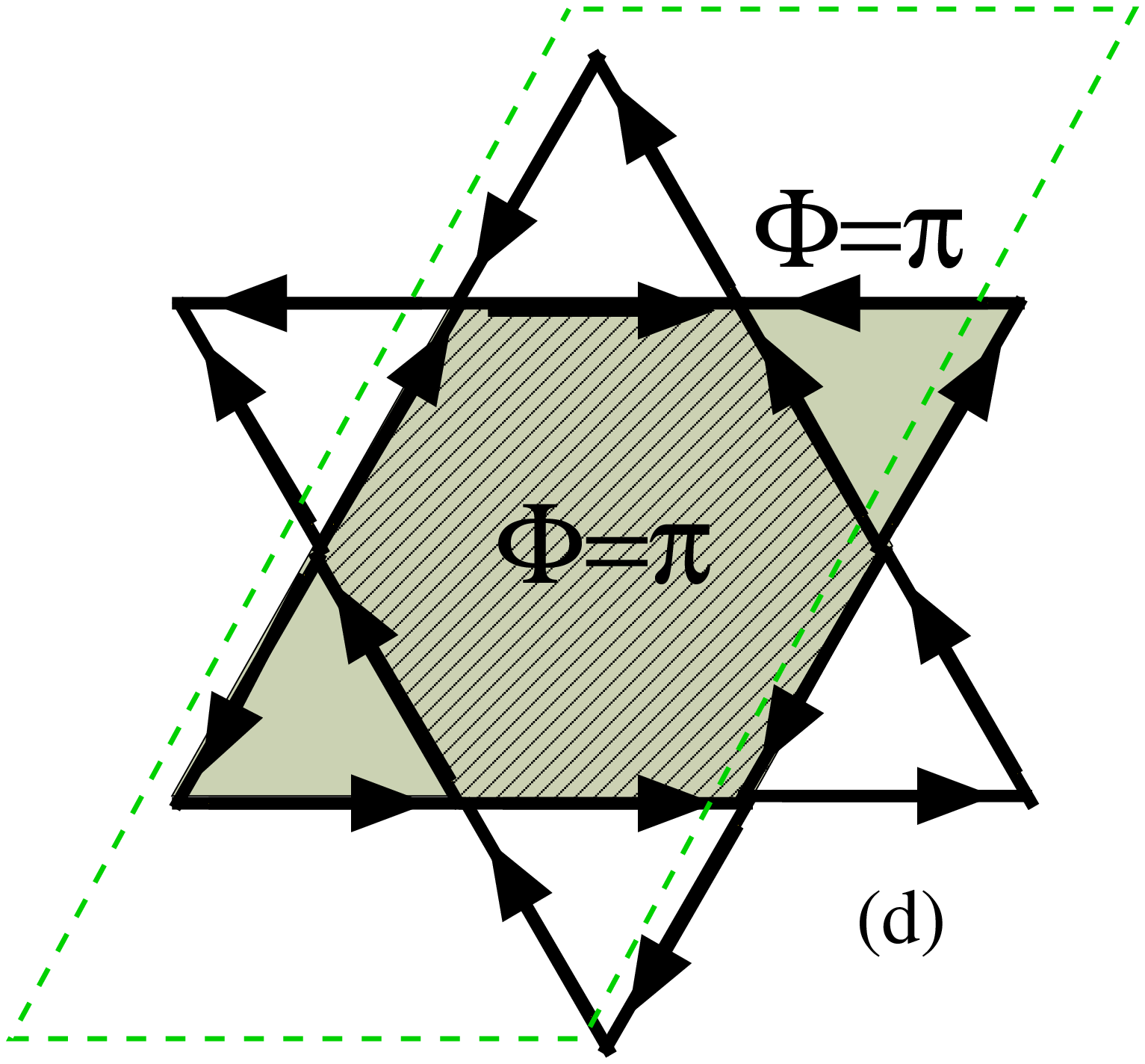,width=3cm,angle=0}
}
\caption{The bond order-parameter ${\cal A}_{ij}$ of the four symmetric \textit{Ans\"atze} of model~Eq.~(\ref{ham}). Its modulus is a constant for all bonds, an arrow
  from $i$ to $j$ means ${\cal A}_{ij}>0$. The fluxes through the hexagon and the rhombus are: in (a) $(0,0)$,  (b) $(\pi,0)$,  (c) $(0,\pi)$  and (d) $(\pi,\pi)$.  The unit-cell is shown by dashed lines (twice larger for the last two \textit{Ans\"atze} ). }
\label{ansatz}
\end{figure}

\noindent
The Schwinger boson representation is written
\begin{eqnarray}
\mathbf{S}_{i}=  \frac{1}{2}
\left(
\begin{array}{cc}a_i^\dag& b_i^\dag\end{array}
\right)
\mathbf{\sigma}
\left(
\begin{array}{c}
a_i \\
b_i\end{array}
\right)
 \hspace{.5cm } n_i \equiv a^\dag_{i} a_{i} +b^\dag_{i}b_{i} =2S,
\label{constraint}
\end{eqnarray}
where $a_i^\dag$ (resp. $b_i^\dag$) creates a boson on site $i$ with spin $\uparrow$ (resp. $\downarrow$), and $\mathbf{\sigma}$ are the Pauli matrices.
To fix
the length of the
 spin, $\mathbf{S}_i^2=(n_i/2)(n_i/2+1)=S(S+1)$, we need to have $n_i=2S$
bosons per site. We define the bond creation operator
\begin{equation}
A_{ij}^{\dagger} \equiv \frac{1}{2}  \left( e^{i \theta_{ij}} a^\dag_{i}
b^\dag_{j} -  e^{-i \theta_{ij}} b^\dag_{i}  a^\dag_{j}    \right),
\label{newA}
\end{equation}
with $\theta_{ij}=D_{ij}/(2J)$.
A similar approach has been developped by Manuel \textit {et al.}\cite{Manuel} on the square lattice.
With this definition and up to small corrections of order $D^2_{ij}/J$, the model takes its standard form:\cite{Auerbach}
\begin{equation}
H =  -2J \sum_{<i,j>}  A_{ij}^{\dagger}A_{ij}   + NzJS^2/2,
\label{sb}
\end{equation}
where $N$ is the number of lattice sites and $z=4$ the coordination number.
Applied to the vacuum of boson, $A_{ij}^{\dagger}$ creates a superposition of a singlet and a triplet state on the $ij$ bond, i.e. the exact ground-state of a single bond of Eq.~(\ref{sb}).
In mean-field theory,\cite{Auerbach,Krish} the quartic Hamiltonian (\ref{sb}) is replaced  by
a self-consistent quadratic Hamiltonian with a bond varying order-parameter ${\cal A}_{ij} \equiv \langle
A_{ij}\rangle$ and the constraint $n_i=2S$ enforced
on average with Lagrange multipliers $\lambda_i$. Up to a constant the mean field Hamiltonian reads:
\begin{eqnarray}
H_{MF} =  &-& 2J
\sum_{<i,j>}  {\cal A}_{ij}^* A_{ij} +  A_{ij}^{\dagger} {\cal A}_{ij} -   |{\cal A}_{ij}|^2 \nonumber  \\  &-&  \sum_{i}  \lambda_i ( n_i -2S ).
\label{sbmf0}
\end{eqnarray}
We now restrict our search to solutions  that do not break the space group symmetry of the Hamiltonian and hence could realize spin-liquid states.
There are only four classes of such states (called \textit{Ans\"atze} in refs.~\cite{Sachdev,Wang} and in the following), labelled by their projective symmetry group,\cite{Wang}  or equivalently by fluxes around hexagons and rhombus
 $(\Phi_{Hex},\Phi_{Rho})=(0,0)$, $(\pi,0)$, $(0,\pi)$, or $(\pi,\pi)$.
The flux $\phi$ around a loop $(i_1, i_2, \dots i_{2n})$ with an even number of links is defined by\cite{Tcherni}
\begin{equation}
 Ke^{i\phi}={\cal A}_{12}(-{\cal A}_{23}^*)\dots {\cal A}_{2n-1,2n}(-{\cal A}_{2n,1}^*).
\end{equation}
It is a gauge invariant quantity.
In each of the four \textit{Ans\"atze}, all ${\cal A}_{ij}$ have the same amplitude $|{\cal A}_{ij}|={\cal A}$ and are real in a well chosen gauge.
Their signs are represented on Fig.~\ref{ansatz}.
These four \textit{Ans\"atze} are identical to those obtained by Wang and Vishwanath for the kagome Heisenberg model: they are fully determined by rotational and translational invariances of the spin Hamiltonian on the lattice.
The $(0,0)$ \textit{Ansatz} corresponds to the $\sqrt{3} \times \sqrt{3}$ state and the $(\pi,0)$ \textit{Ansatz} to the $Q=0$ state, originally found by Sachdev,\cite{Sachdev} while the last two involve larger unit cells and  may as well be relevant for longer range interaction or ring exchange.\cite{Wang,domenge}

Using the translation symmetry of the mean-field \textit{Ans\"atze}, the Hamiltonian is Fourier-transformed,
\begin{equation}
H_{MF} =
\sum_{\mathbf{q}} \phi^\dag_\mathbf{q} N_\mathbf{q} \phi_\mathbf{q} +NJz {\cal A}^2 + (2S+1)N\lambda
\label{sbmf}
\end{equation}
with
\begin{equation}
\phi^\dag_\mathbf{q} \equiv ( (a_{1\mathbf{q}})^\dag,\dots ,(a_{m\mathbf{q}})^{\dag}, b_{1-\mathbf{q}}, \dots ,b_{m-\mathbf{q}})
 \end{equation}
\begin{equation}
a_{ i\mathbf{q}} \equiv\frac{1}{\sqrt{N/m}}\sum_{\mathbf{x}} e^{-i\mathbf{q}\mathbf{x}} a_{i \mathbf{x} },
 \end{equation}
(the same for the $b$ operators), $m$ is the number of
sites in the unit-cell: $m=3$ for the first two \textit{Ans\"atze} ($N_\mathbf{q}$ is $6 \times 6$), $m=6$ for the second two. A given site is defined by a sublattice $i\in[1,m]$ and unit-cell $\mathbf{x}$ indices.
The
Hamiltonian is diagonalized using a
numerically-constructed\cite{Colpa} Bogoliubov transformation
\begin{equation}
 \mathbf{\phi}_{\mathbf{q}} =  P_{\mathbf{q}}\tilde{\phi}_{\mathbf{q}}
\hspace{1cm}
  P_{\mathbf{q}}=\left(\begin{array}{cc}
U_{\mathbf{q}}&-V_{\mathbf{q}} \\
V_{\mathbf{q}}&U_{\mathbf{q}}
\end{array}\right)
\label{Bogoliubov}
\end{equation}
where $U_{\mathbf{q}}$ and $V_{\mathbf{q}}$ are $m\times m$ matrices.
\begin{equation}
H_{MF} =
\sum_{\mathbf{q} \mu } \omega_{\mathbf{q}\mu} \tilde{\phi}^{\dagger}_{\mathbf{q} \mu}  \tilde{\phi}_{\mathbf{q} \mu } + NJz {\cal A}^2+(2S+1)N\lambda
\label{eq:hamdiag}
\end{equation}
where $\omega_{\mathbf{q} \mu}$ is the dispersion relation of the $\mu=1,...,2m$ spinon modes.
Each mode is twice degenerate because of the time-reversal symmetry.
The ground-state $\left|\tilde 0\right>$ is  the vacuum of the Bogoliubov bosons.
At zero temperature, ${\cal A}$ and $\lambda$ are determined by
extremizing the total energy, subject to the constraints:
\begin{equation}
{\cal A}=|\langle A_{ij}\rangle|
\hspace{1cm} \langle n_i \rangle=2S
\label{constraints}
\end{equation}
(the energy is in fact a saddle point, minimum in ${\cal A}$  and maximum in $\lambda$\cite{yavors'kii}).

\section{Phase diagram}
\label{Section:phasediagram}
\begin{figure}
\begin{center}
 \includegraphics[width=4cm]{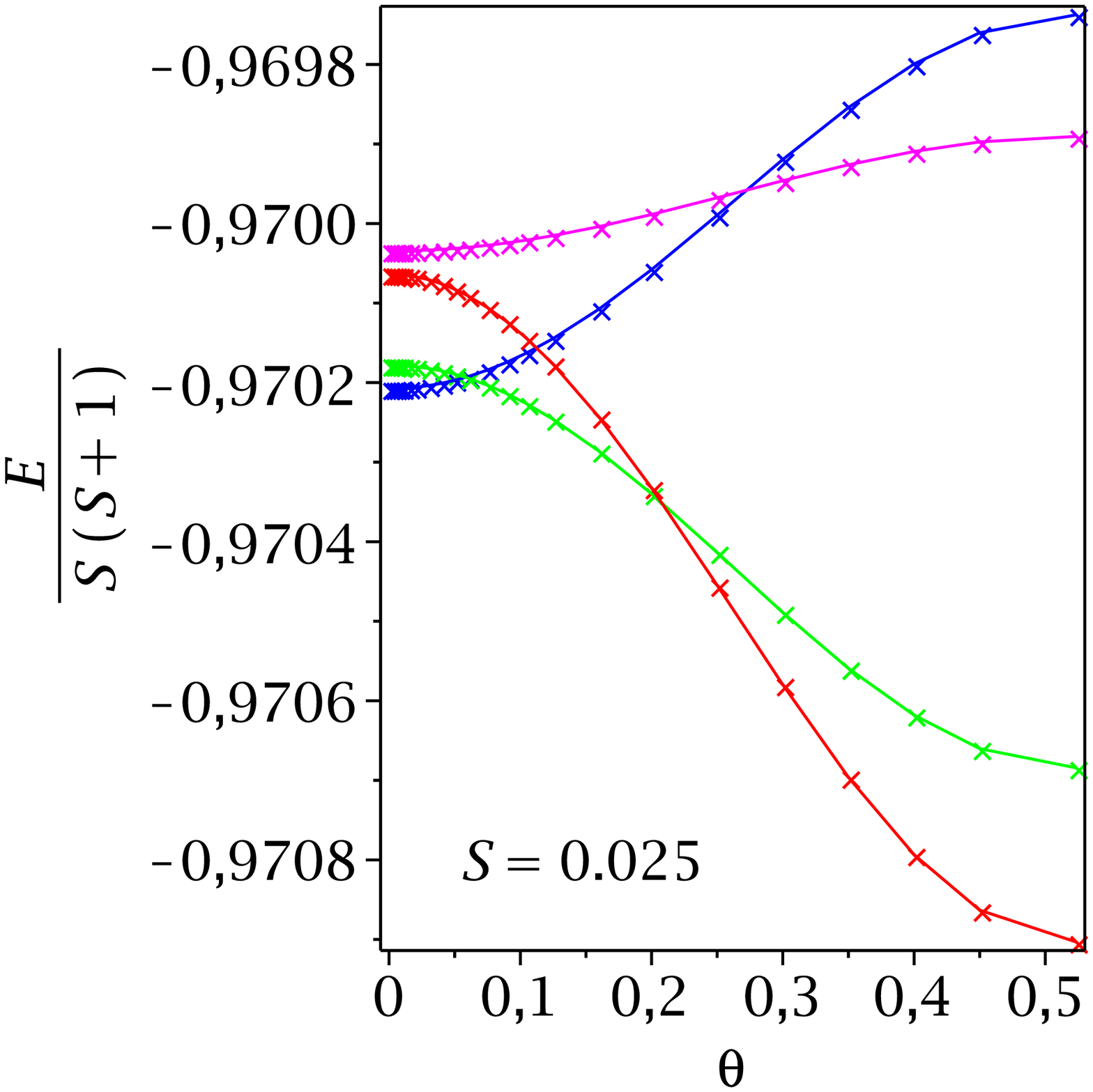}
 \includegraphics[width=4cm]{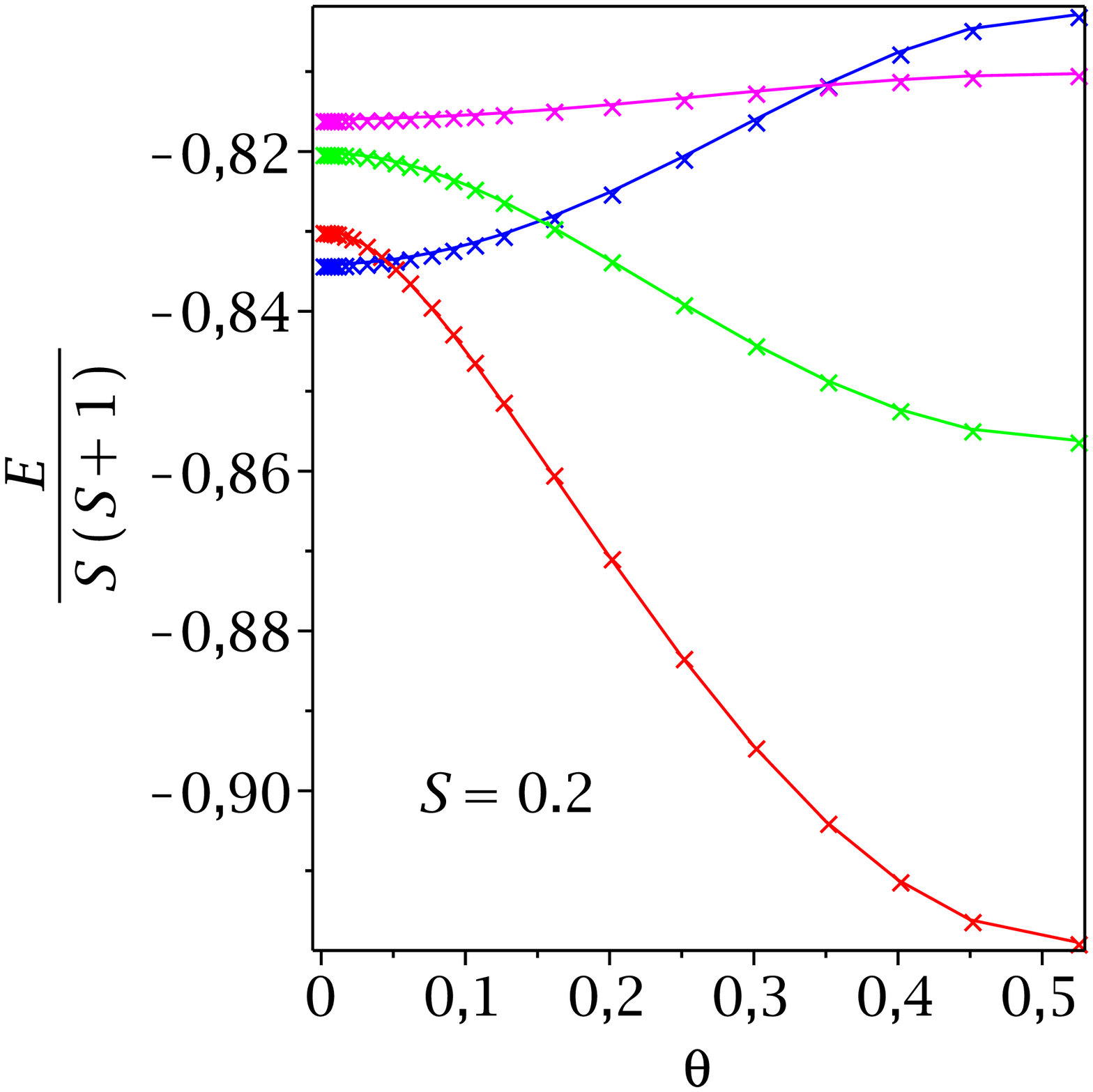}
 \includegraphics[width=4cm]{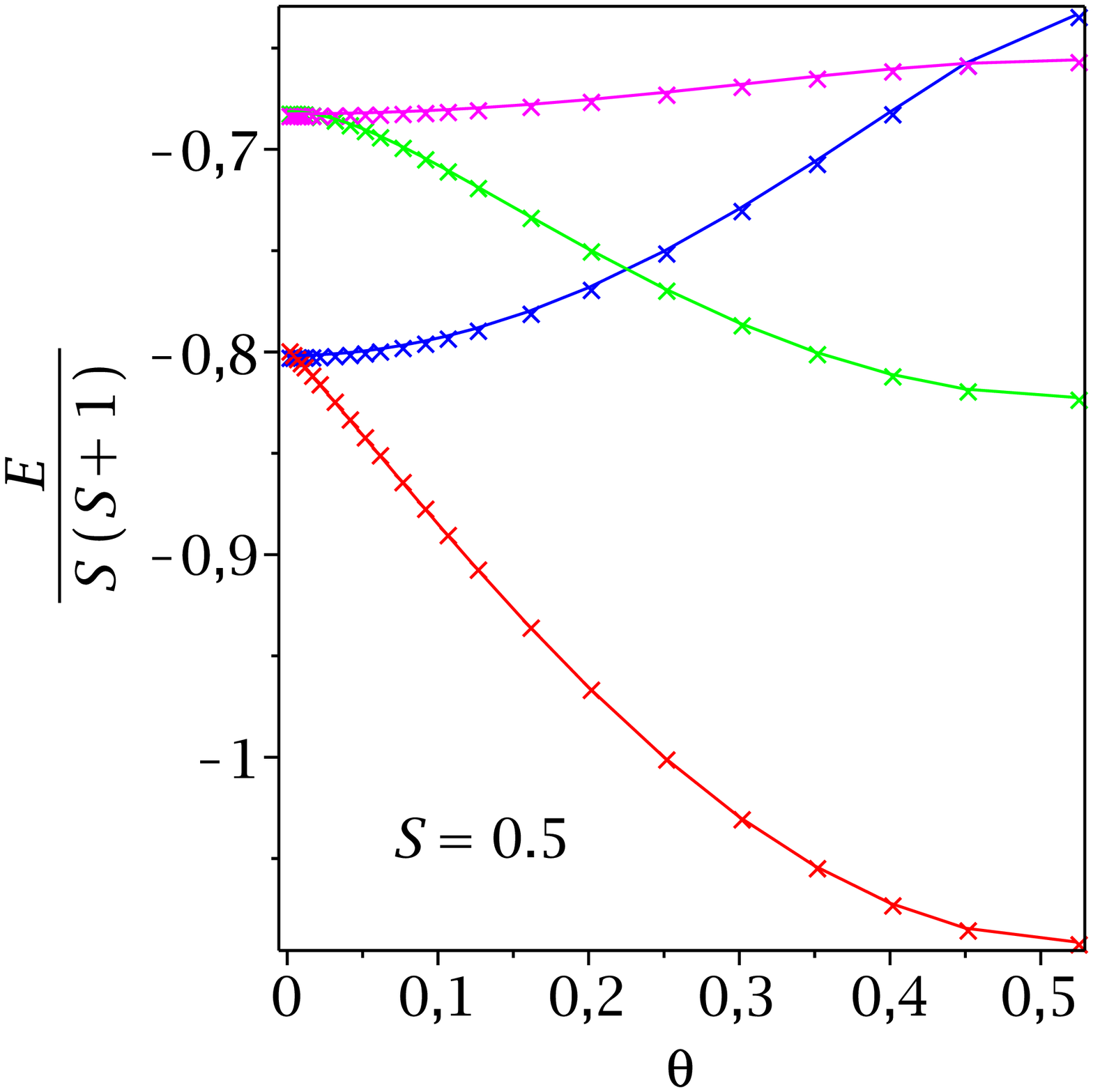}
\caption{(Color online) Ground state energies of the four \textit{Ans\"atze} (blue=$(0,0)$, red=$(\pi,0)$, green=$(0,\pi)$, magenta=$(\pi,\pi)$)  vs. $\theta=D/2J$, for $S=0.025$, $0.2$, $1/2$ (top to bottom). }
\label{fig:Espins}
\end{center}
\end{figure}

\begin{figure}
\begin{center}
 \includegraphics[width=6.5cm]{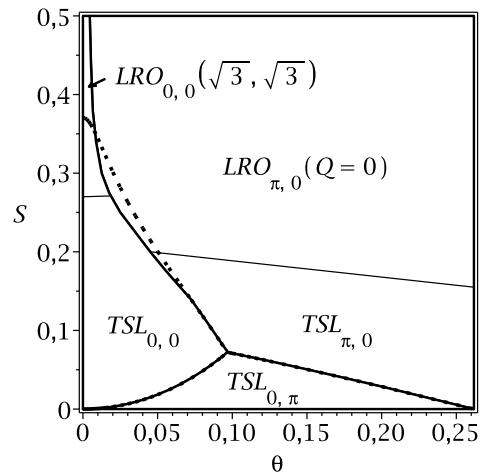}
\caption{Phase diagram at zero temperature (spin $S,\theta=D/(2J)$): topological spin liquid (TSL) and N\'eel long-range ordered (LRO) phases characterized by their fluxes through hexagons and rhombi.  For larger $S$ the region of existence of the $(0,0)$ phase shrinks. Dashed lines are the result of a perturbative expansion at small $S$ (see text). }
\label{fig:diagramme}
\end{center}
\end{figure}

To obtain the phase diagram, the two self-consistent equations (\ref{constraints}) are implemented numerically
for each of the four \textit{Ans\"atze} of Fig.~\ref{ansatz}. The energies of the different \textit{Ans\"atze} are shown in Fig.~\ref{fig:Espins} versus $\theta=D/(2J)$, for  three values of $S$:  $0.025$, $0.2$ and  $1/2$.
The corresponding full phase diagram  of the model is shown in Fig.~\ref{fig:diagramme}.

Before discussing the predictions of this model, let us remark that in
the small $S$ limit all these phases can in fact be captured by an
analytic perturbative expansion in term of flux through closed
loops.\cite{Tcherni} At small $S$, the density of bosons is small and
the constraints (\ref{constraints}) imply that ${\cal A}_{ij}$ must be
small compared to $\lambda$. The mean-field energy can then be
expanded in terms of gauge-invariant products of bond order parameters
$ {\cal A}_{ij} $ along closed loops.  Following Tchernyshyov \textit {et al.},~\cite{Tcherni} we have computed the expansion up to loops of length 16 
in order to calculate
the energy difference between the four \textit{Ans\"atze}. The results
of these calculations give the low $S$ phase boundaries (dashed lines
superimposed to the exact results in Fig.~\ref{fig:diagramme}). It is
seen on this example that the so-called flux expulsion conjecture\cite{Tcherni} which
predicts that the ground-state in non frustrated models has zero flux
through any closed loop does not apply to frustrated
problem, where the $(0,\pi)$ and $(\pi,0)$ appear as ground-states in
an extended range of parameters.

For the sake of clarity, we will discuss the full phase diagram (Fig.~\ref{fig:diagramme}) and postpone to the next paragraph the illustration on the spinon spectrum of the differences between topological spin liquids and N\'eel ordered phases.
For $S=1/2$, there is a direct first-order transition between the long-ranged N\'eel ordered $\sqrt{3}\times\sqrt{3}$ and
$Q=0$ phases for a finite \DM coupling. This finite range of existence of the $\sqrt{3}\times\sqrt{3}$ phase shrinks with increasing values of the spin, which is fully compatible with the classical solution.~\cite{Elhajal}

For low $S$ values, and increasing $\theta$, Figs.~\ref{fig:Espins}
and \ref{fig:diagramme} show a sequence of first-order transitions
between the $(0,0)$ ($\sqrt{3} \times \sqrt{3}$ short-range
fluctuations), $(0,\pi)$, and $(\pi,0)$ ($Q=0$, short range
fluctuations) spin liquid phases. (The ($\pi,\pi$) state is always at
higher energy and never realized.)  The $(0,\pi)$ state was argued to
be stabilized by four-spin interactions up to a large critical
$S$.\cite{Wang} It also appears here in a small part of the phase
diagram for very small $S$ but first-order transitions prevent its
stability at larger $S$.

In the absence of \DM coupling, the $S=1/2$ results of this approach
are qualitatively not consistent with exact diagonalisations which
point to a non magnetic phase.  But in the range of parameters around
$S \sim 0.2$, the $(\pi,0)$ Schwinger boson mean-field results are
qualitatively not very far from exact diagonalization results: there
are short-ranged $Q=0$ correlations in the Heisenberg
case\cite{laeuchli,Jiang} and a second-order phase transition to
120$^o$ $Q=0$ N\'eel order under the effect of \DM
coupling.\cite{Cepas}

As already mentioned, it may be that a theory beyond mean-field, with a better treatment of the constraint  shifts the region $S\sim 0.2$ to the physically accessible $S=1/2$.  Indeed in the Schwinger bosons mean-field approach it is well known~\cite{Auerbach} that there are large fluctuations of the number of bosons. As a consequence, the square of the spin operator
\begin{equation}
\langle
\textbf{S}_i^2\rangle=S(S+1)+ (\langle n_i^2 \rangle -\langle n_i\rangle^2)/4
\label{S2}
\end{equation}
takes a value 3/2 times larger than $S(S+1)$ (at
$D=0$).\cite{noteWick} The prefactor is even larger in the presence of
\DM interaction and amounts to $\sim 1.75$ for $\theta=0.25$.
From the physical spin-1/2 point of view, the mean field approximation
leads to (unwanted) extra fluctuations and, on
average, the spin length is larger than assumed (because of (\ref{S2})).
The SU(2) symmetry of the Heisenberg model can be generalized to Sp$({\cal
N})$ (which reduces to SU(2) when ${\cal N}=1$)
by duplicating $\cal N$ times each pair $(a_i,b_i)$ of boson operators:
$(a_{i \alpha},b_{i \alpha})$ where $\alpha=1\cdots{\cal N}$ is a
``flavour'' index.\cite{ReadSachdev91}
It can be shown that the different boson flavors decouple in the limit
of large ${\cal N}$, leading to ${\cal N}$ uncorrelated copies
of the single flavor problem, for which the {\it exact} solution is
given by the present {\it mean-field} treatment of the ${\cal N}=1$
model. Thus, it is only in the ${\cal N}=\infty$
limit of the model that the mean field treatment becomes exact and that
the fixed ``spin length'' is recovered. However, the mean-field theory can describe qualitatively the
magnetically ordered phases and the deconfined  $\mathbb Z_2$ spin
liquid phases of the SU(2) model.\cite{Gregoire}

\section{Spinon spectrum and quantum phase transition from a topological spin liquid to a N\'eel ordered phase}
\label{Section:spinon}

\begin{figure}
\begin{center}
 \includegraphics[width=4cm]{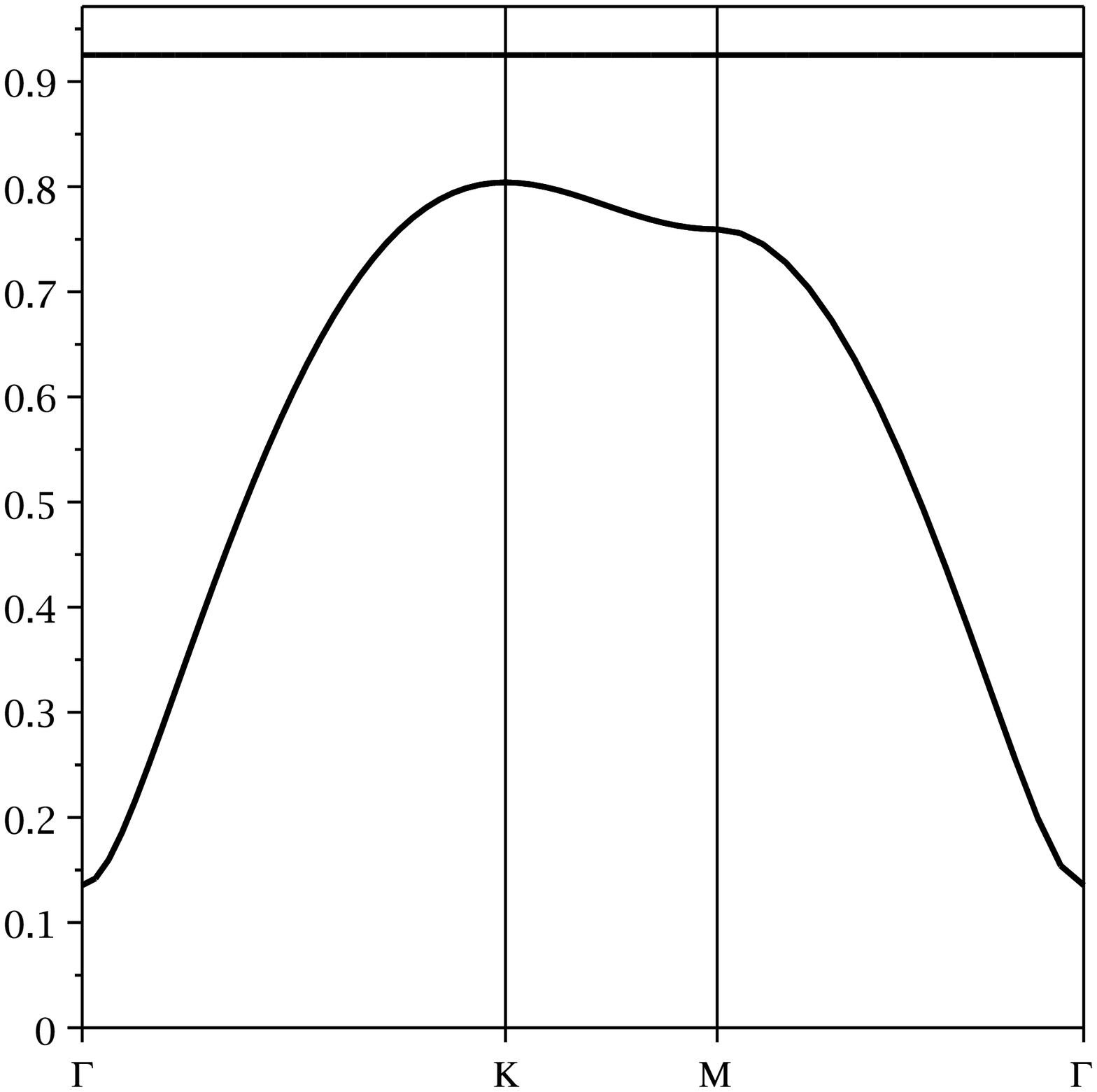}
 \includegraphics[width=4cm]{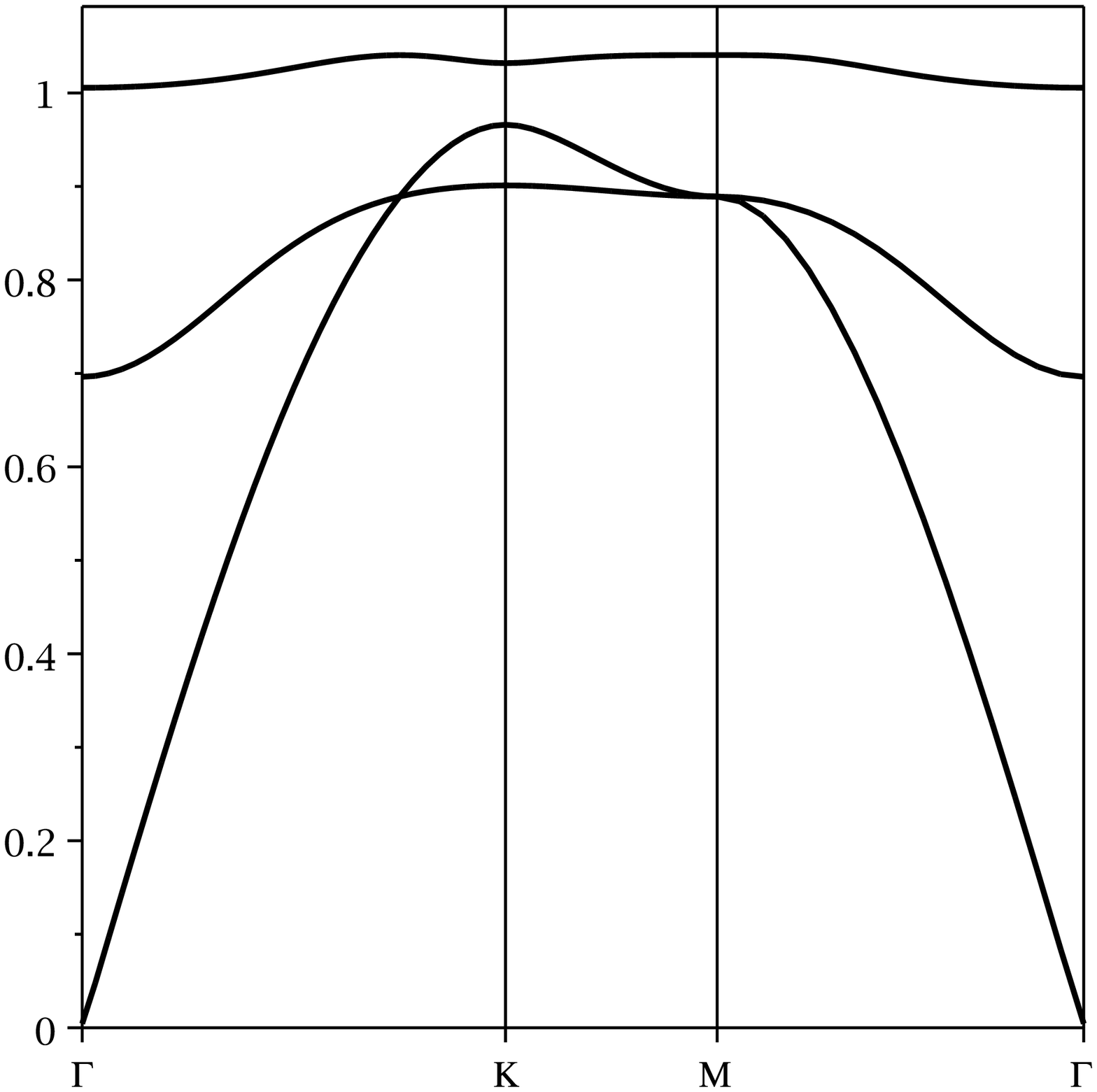}
\caption{Spinon dispersions along $\Gamma-K-M-\Gamma$ of the first Brillouin Zone (see Fig.~\ref{stat} for the definition) for the  $(\pi,0)$  \textit{Ansatz},  $S=0.2$. Left: $\theta = 0$, the system   has a gap and is a topological spin liquid with short range $Q=0$ correlations. Right: $\theta = 0.25$, the lower branch becomes gapless at $\Gamma$ in the thermodynamic limit and gives rise to the  Goldstone mode of the long-range  N\'eel order.}
\label{fig:Espinons}
\end{center}
\end{figure}

\textit{Spin liquid phases (low $S$).}  In the low $S$ regime,  the spinon spectrum  of Eq.~(\ref{eq:hamdiag}) is gapped everywhere. Fig.~\ref{fig:Espinons} (left) gives a typical example of such a  spectrum for the \textit{Ansatz} $(\pi,0)$. The spinon band structure is shown in the first Brillouin zone, it has a gap of order ${\cal O}(1)$ at point $\Gamma$ indicative of short range $Q=0$ correlations. This gap does not go to zero in the thermodynamic limit: this phase is a disordered topological spin liquid.  Adding a \DM perturbation has two effects on the spectrum. It lifts the degeneracy of the lower band corresponding to the symmetry breaking of the model from $SU(2)$ to $U(1)$
 and it reduces the gap of the spinon mode that has the chirality opposite to that of the \DM field.

\textit{Bose-Einstein condensation.}  With increasing $S$ or $\theta$, the gap decreases and above a critical spin $S_c(\theta)$, the spinon spectrum shows a finite-size gap, which collapses with system size $N$ as  ${\cal O}(1/N)$. Such a spectrum is shown in Fig.~\ref{fig:Espinons} (right).
In the thermodynamic limit, the bosons condense in the soft mode  (noted $\tilde{\phi}_{\mathbf{q}_0 l_0 \sigma}$
with $\sigma=\uparrow,\downarrow$).\cite{simplicity}  This  gives a macroscopic contribution to the total number of Schwinger bosons:
\begin{equation}
\frac{1}{2S}\sum_i \langle n_i \rangle
=N
=\sum_{\mathbf{q}ij}\frac{ |V_{\mathbf{q} ij}|^2}{S}
=x_{\mathbf{q_0}} N + \sum_{\mathbf{q}\neq\mathbf{q_0},ij}\frac{ |V_{\mathbf{q} ij}|^2}{S}
\label{BEC}
\end{equation}
The condensed fraction  $x_{\mathbf{q_0}}$ in the soft mode is of order ${\cal O}(1)$ (or equivalently $|V_{\mathbf{q_0} i l_0}| \sim \sqrt{N}$).
 The transition to this Bose-Einstein condensed phase corresponds to the development of long-range antiferromagnetic  correlations, as can be seen by computing the static structure factor:
\begin{equation}
S^{xx}(\mathbf{Q})= \frac{3}{4N} \sum_{i,j} e^{-i\mathbf{Q}.(\mathbf{R}_i-\mathbf{R}_j)} \langle \tilde{0} |  S_i^x S_j^x |\tilde{0} \rangle
\label{staticstructurefactor}
\end{equation}
where $\mathbf{R}_i$ is the position of site $i$ and $x$ is an axis in
the easy plane perpendicular to $\mathbf{D_{ij}}$.

\begin{figure}[htbp]
\centerline{
\psfig{file=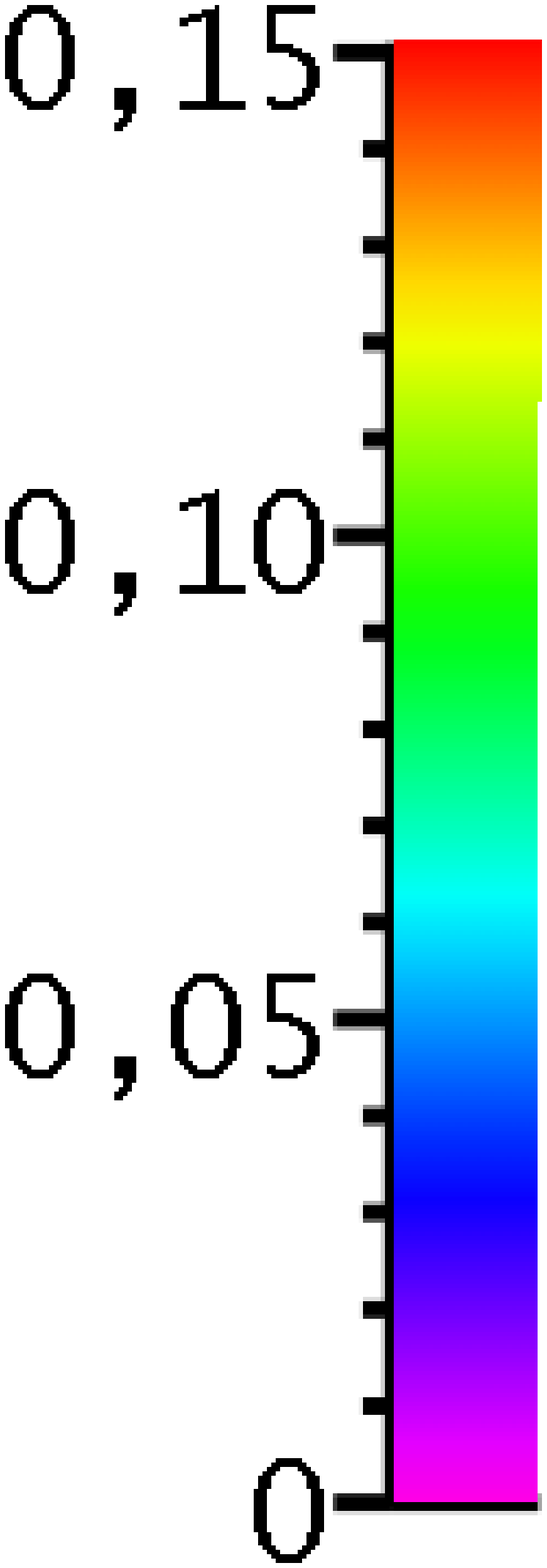,width=0.8cm,clip,trim=0.65cm -2cm 1.8cm 0cm}
\psfig{file=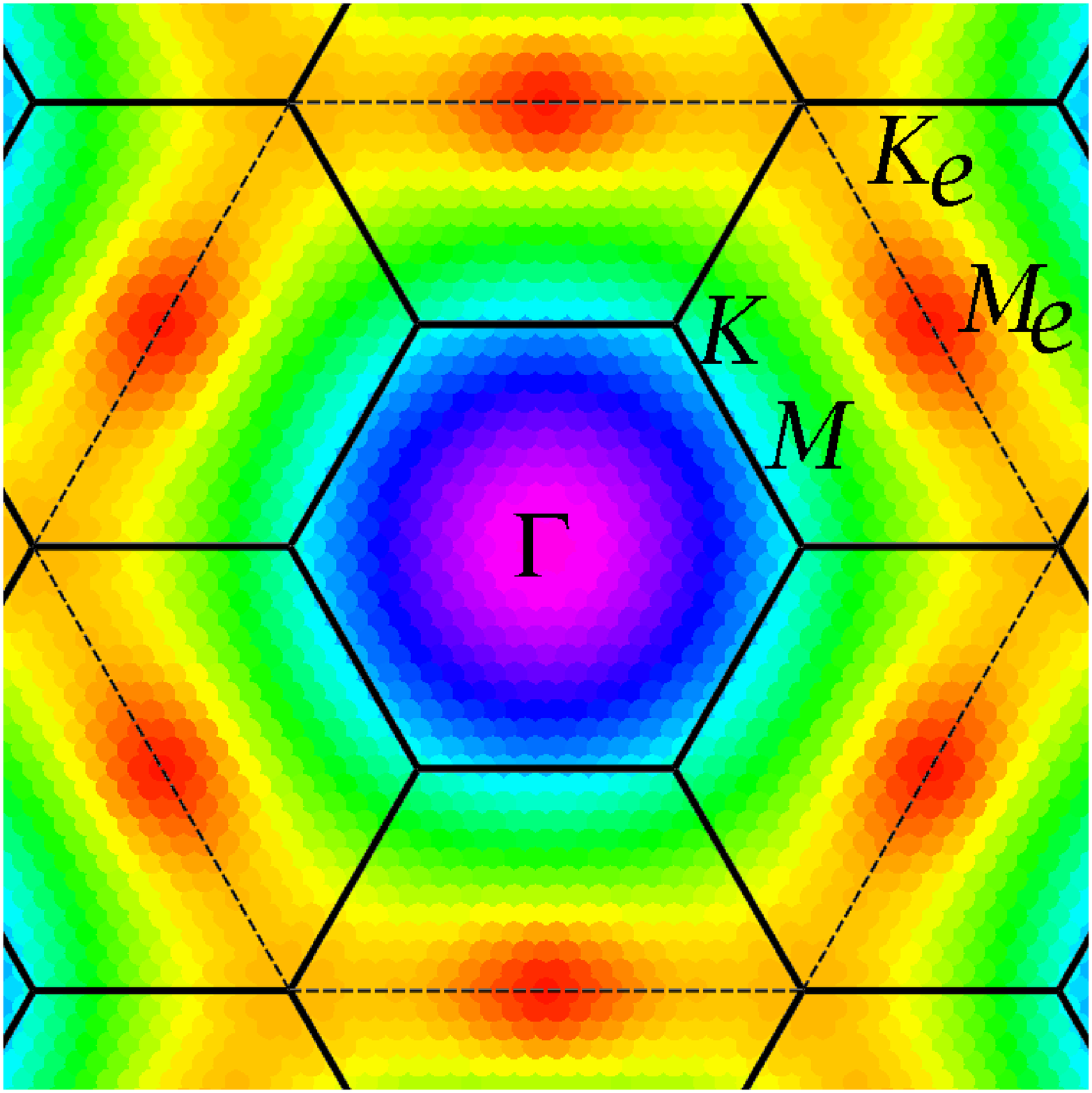,width=3.2cm,trim=0.4cm 0cm 0.9cm 0cm}
\psfig{file=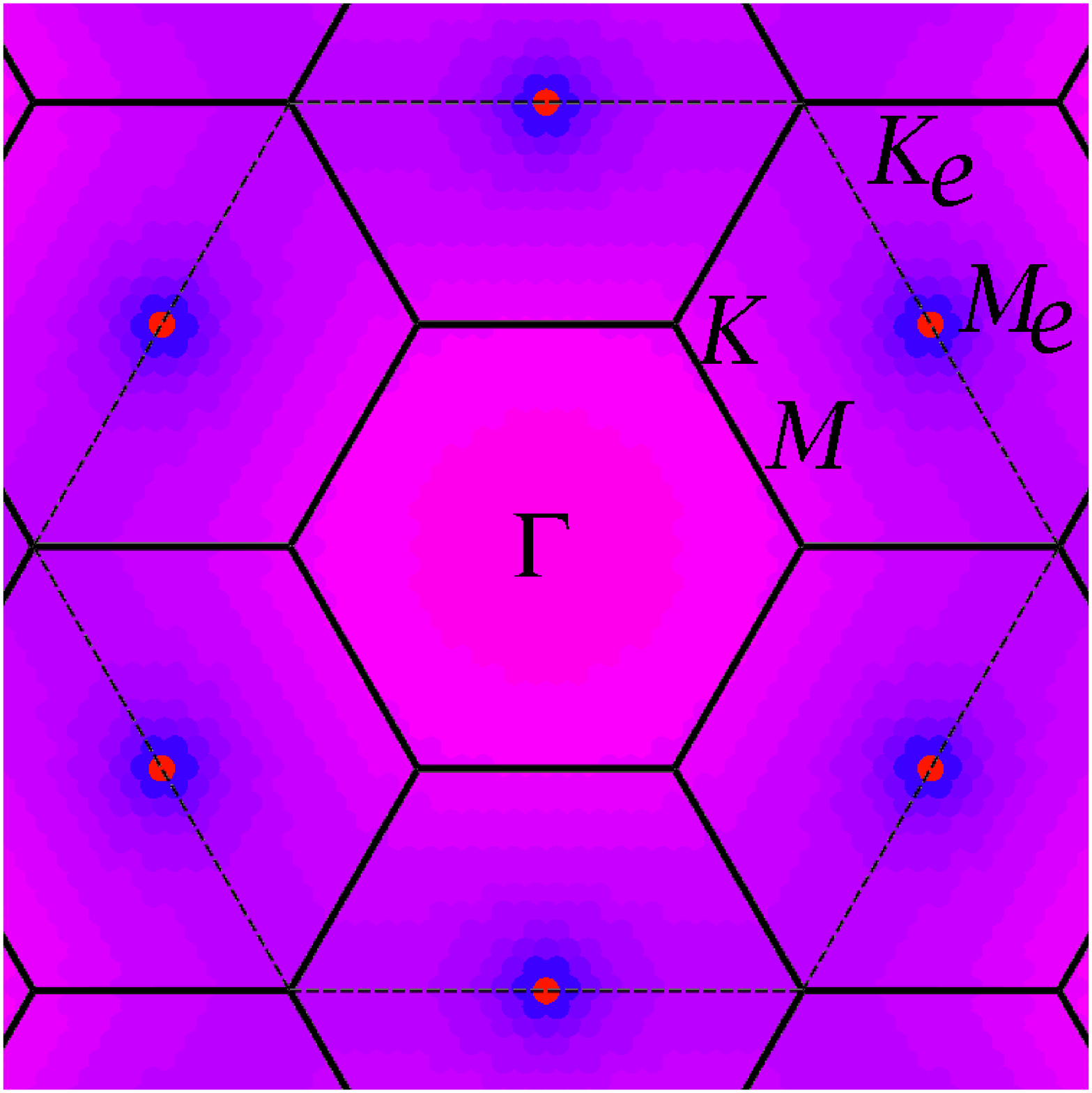,width=3.2cm,trim=0.1cm 0cm 1.2cm 0cm}
\psfig{file=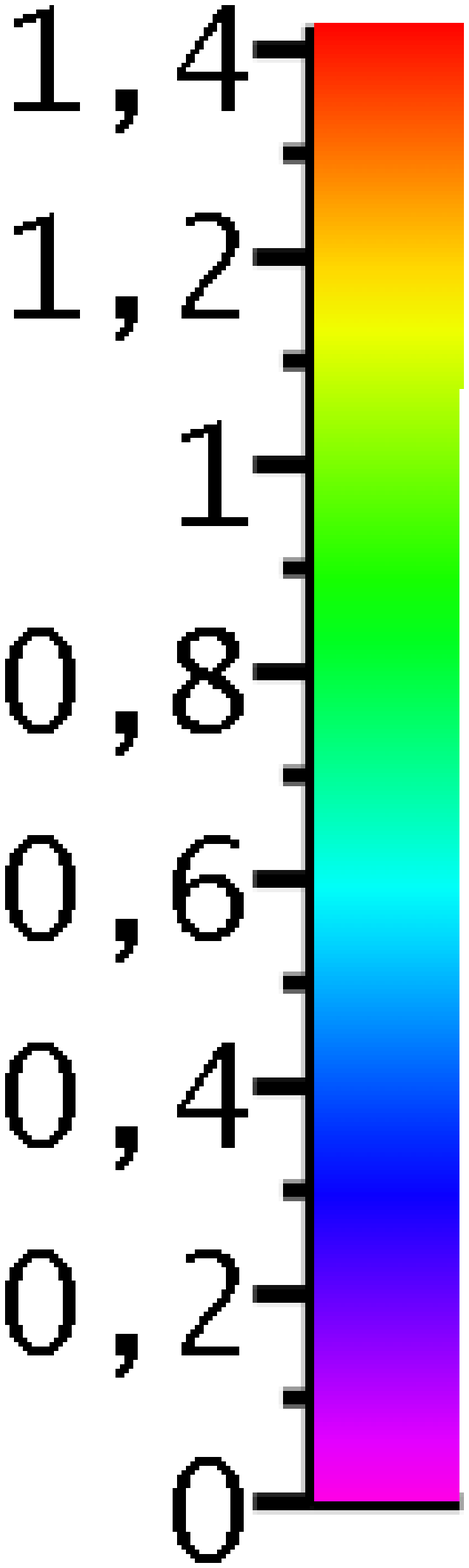,width=0.8cm,trim=0.65cm -2cm 1.8cm 0cm,clip}}
\caption{(Color online) Static structure factor $S^{xx}$ in the extended Brillouin zone for the $(\pi,0)$ state ($Q=0$) for a lattice size $N=1296$. In the spin liquid phase $TSL_{\pi,0}$ (left, $S=0.2$, $\theta=0$) there are broad features about $\mathbf{M}_e$ which become peaks with divergent intensity in the ordered phase $LRO_{\pi,0}$ (right,  $S=0.2$, $\theta=0.25$).  }
\label{stat}
\end{figure}

\begin{figure}
\begin{center}
\includegraphics[width=7cm]{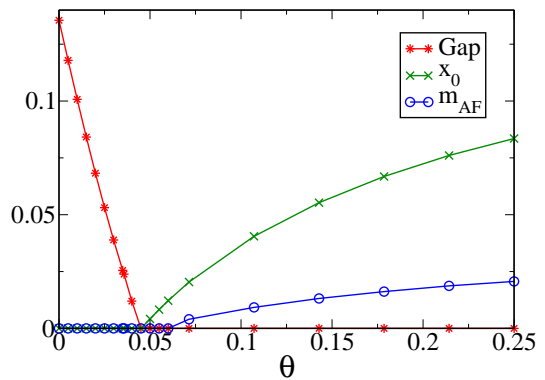}
\caption{Gap of the soft mode, order-parameter $m_{AF}$ and condensed fraction, $x_{\mathbf{q}_0}$ extrapolated to the thermodynamic limit for the $(\pi,0)$ \textit{Ansatz} as a function of $\theta=D/(2J)$ ($S=0.2$).}
\label{fig:limit_thermo}
\end{center}
\end{figure}

The  difference in static structure factor between topological spin liquid and N\'eel order phase is illustrated in Fig.~\ref{stat}, for the  $(\pi,0)$ \textit{Ansatz} across the Bose-Einstein
condensation.
In the spin liquid phase, the structure factor has broad features located at $\mathbf{Q}= \mathbf{M}_e$ (Fig.~\ref{stat}, left)
and at equivalent reciprocal points (these are the $\Gamma$ points of the second Brillouin zone).
This structure factor looks very similar to exact diagonalization and DMRG results.\cite{laeuchli,Jiang}  The features become sharp in the Bose-Einstein condensed
phase (Fig.~\ref{stat}, right), where $S^{xx}(\mathbf{M}_e)$ becomes proportional to $N$:
\begin{equation}
\frac{3}{4N}S^{xx}(\mathbf{M}_e) = m_{AF}^2 + \frac{C^{st}}{\sqrt{N}} + \cdots
\label{fsc}
\end{equation}
where $m_{AF}^2$ is the order-parameter corresponding to long-range
correlations of the $120^o$ $Q=0$ N\'eel type.\cite{normalization} We
have extracted $m_{AF}^2$ by fitting the numerical results (up to
$N=1764$) to Eq.~(\ref{fsc}) with finite-size corrections up to order
$1/N$.\cite{fss} The extrapolation to the thermodynamic limit of
$m_{AF}$, together with the condensate fraction $x_{\mathbf{q}_0}$ and
the gap of the soft spinon are given in Fig.~\ref{fig:limit_thermo}
for $S=0.2$. While near the second-order phase transition, the
extrapolation of the condensed fraction behaves very smoothly and
vanishes right at the point where the gap opens, the extrapolation of
the order-parameter gives a small shift.  We note that $m_{AF}$ is
very small (a few percent) in this range of parameters, and,
therefore, more accurate extrapolations would require using larger
system sizes close to the critical point.  For strong enough N\'eel
order, however, the two order-parameters are clearly proportional
($m_{AF} \propto x_{\mathbf{q_0}}$). 

\begin{figure}
\begin{center}
\includegraphics[width=7cm,trim=2cm 2cm 2cm 1cm,clip]{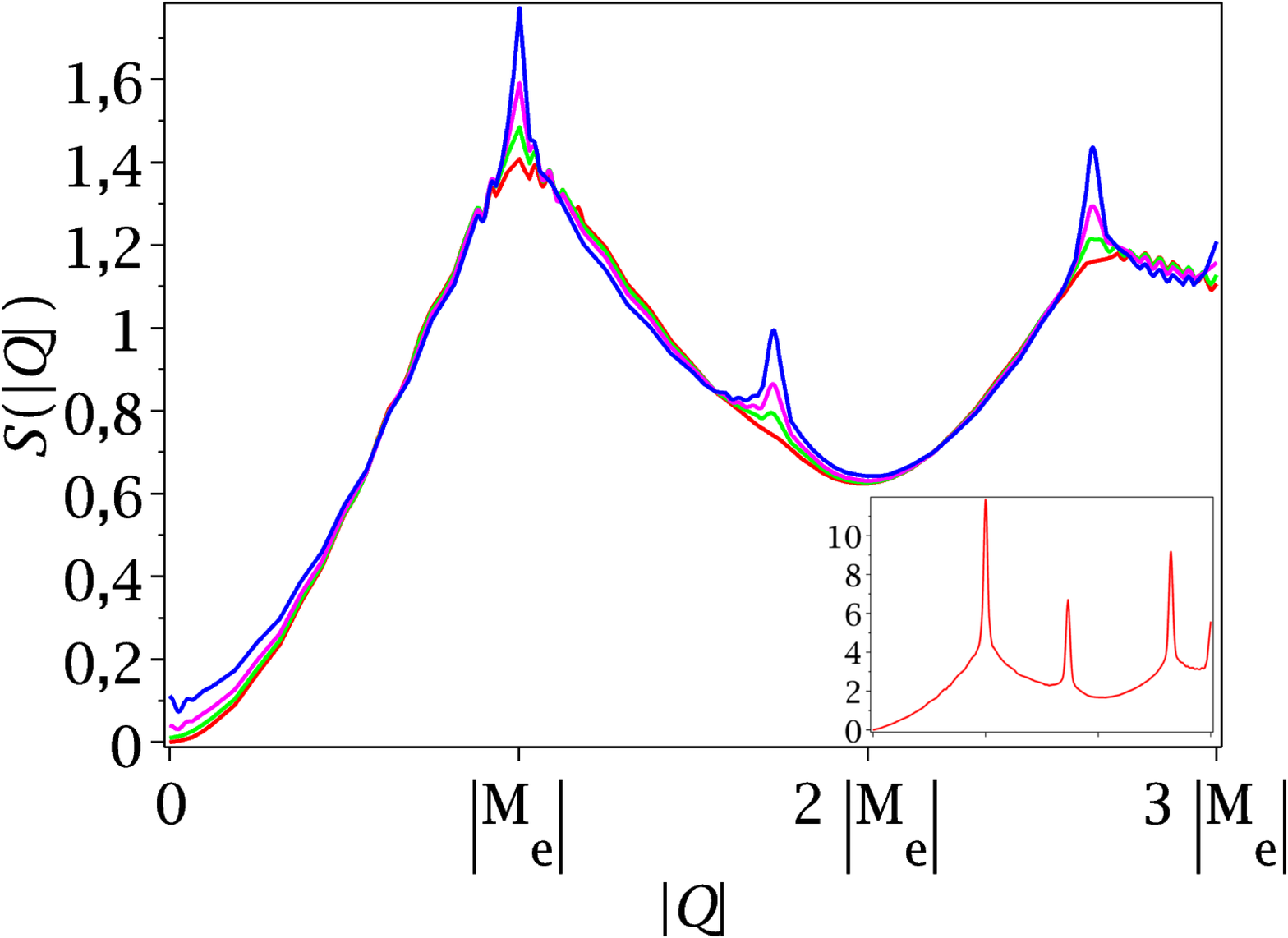}
\caption{(Color online) Static structure factor (powder-averaged) across the quantum phase transition from the spin-liquid to the N\'eel phase ($S=0.2$ and $\theta=0$ (red), $0.05$ (green), $0.1$ (magenta) and $0.2$ (blue)). Inset: $S=1/2$, $\theta=0$. }
\label{powder}
\end{center}
\end{figure}

We have also calculated the static structure factor for powder samples
(denoted by $S(|\mathbf{Q}|)$) by averaging
(\ref{staticstructurefactor}) over all directions of $\mathbf{Q}$. In
the $(\pi,0)$ spin-liquid phase (Fig.~\ref{powder} at small $D$), the
overall shape is characteristic of short-range correlations of
liquids.  We can compare with the calculation of the diffuse
scattering for the classical spin-liquid kagom\'e antiferromagnet by
Monte Carlo simulation.\cite{Reimers} Here the position of the first
broad peak is at $|\mathbf{M}_e|$ instead of $|\mathbf{K}_e|$ (and the second  broad feature is at $\sqrt{7}|\mathbf{M}_e|$). This
simply reflects the difference of short-range correlations of the
$(\pi,0)$ \textit{Ansatz} and the $\sqrt{3} \times \sqrt{3}$ classical
spin-liquid.  In addition, compared with classical Monte Carlo
simulations, we find no intermediate shoulder between the two main
broad peaks (except for a little hump at  $\sqrt{3}|\mathbf{M}_e|$), a point which seems in fact to be closer to recent
experiments on a spin-liquid deuteronium jarosite.\cite{Fak}
Moreover, since the response is due to quantum fluctuations we expect a rather weak sensitivity to the temperature up to temperatures of the order of a fraction of $J$. 
When $D$ increases, we see the development of the
Bragg peaks in the ordered phase, which increase as the square of the order parameter when we go
deep into the ordered phase (Fig.~\ref{powder} (inset)). In the
ordered phase we can identify two distinct contributions to
(\ref{staticstructurefactor}) by using the sum rule
\begin{equation}
S^{xx}(\mathbf{Q})=\frac{1}{2\pi}\int d\omega S^{xx}(\mathbf{Q}, \omega)
\end{equation}
There are the Bragg peaks ($\omega=0$) and also the inelastic collective modes (which we will detail below) which give the additional magnetic background scattering (which is the only contribution to scattering in the spin liquid phase). It is noteworthy that the latter is relatively strong for low spin (Fig.~\ref{powder}) and becomes relatively much smaller once the order-parameter is large (inset of Fig.~\ref{powder}). In fact the transfer of spectral weight from the magnetic continuum background to the Bragg peaks goes as the square of the order parameter.  Note also that $S(|\mathbf{Q}|)$ does not vanish any more for small $\mathbf{Q}$ at $D \neq 0$, this is because in the presence of the anisotropy the ground state is no longer an SU(2) singlet. Although the effect is small the measurement at small $|\mathbf{Q}|$ in the spin-liquid phase could help to figure out what the anisotropy is (or give an upper bound when the signal is small, see, e.g. Ref.~\onlinecite{Lee}).
In the ordered phase, the finite uniform susceptibility should give a finite contribution at $\mathbf{Q}=0$ but we recall that this contribution is absent for the U(1) singlet ground state of  Schwinger boson theory.

\section{Dynamical spin structure factor }
\label{Section:dynamical}
The Schwinger boson approach allows to calculate the dynamical response of the system, which is interesting both theoretically and
for a direct comparison with experiments. The inelastic neutron
cross-section is proportional to the spin dynamical structure factor
\begin{eqnarray}
S^{\alpha \alpha}(\mathbf{Q},\omega)
&=& \int^{+\infty}_{-\infty} dt e^{i\omega t} \langle  S_{\mathbf{Q}}^{\alpha}(t) S_{\mathbf{-Q}}^{\alpha}(0)\rangle \\
&=&2\pi \sum_{p} |\langle \tilde 0 |S_{\mathbf{Q}}^{\alpha}|\tilde p \rangle|^2  \delta(\omega-\omega_{p})
\label{eq:struct1pm}
\end{eqnarray}
where $\alpha=x,y,z$ depending on the polarization of the incident neutrons,
$|\tilde 0 \rangle$ is the ground-state and
\begin{equation}
\mathbf{S}_{\mathbf{Q}}
=\sqrt{\frac{3}{4N}}\sum_{\mathbf{x} i} e^{-i\mathbf{Q}.\mathbf{R}_{i\mathbf x}}\mathbf{S}_{i\mathbf{x}},
\label{fouriert}
\end{equation}
with $\mathbf{R}_{i\mathbf x}$ the position of the site $i\mathbf x$.
We use the Fourier transform and the Bogoliubov transformation to
express $\mathbf{S}_\mathbf{Q}$ in terms of quasiparticle operators
(\ref{Bogoliubov}). At zero temperature, since $|\tilde 0\rangle$ is
the vacuum of quasiparticles, only creation operators are retained. Given that $\mathbf{S}_{i\mathbf{x}}$ is quadratic
in boson operators, we can only create spinons by pair.
For example, the following term is present in $S^z_\mathbf Q$
\begin{eqnarray}
\sum_{\mathbf{q}ln}
-V_{(\mathbf{Q}+\mathbf{q})il}^* U_{\mathbf{q} in}
 \tilde b_{l-(\mathbf{Q}+\mathbf{q})} \tilde a_{n\mathbf q},
\label{2spinons}
\end{eqnarray}
which applied to the left of the matrix element (\ref{eq:struct1pm}),
 creates a pair of spinons $|\tilde p\rangle$ with energy $\omega_p = \omega_{(\mathbf{Q}+\mathbf{q}) l} + \omega_{-\mathbf{q} n}$ and wave-vector $\mathbf{Q}$.
 The intensity of the transition is obtained by computing the product of matrices, such as in (\ref{2spinons}), for each pair of modes.

\begin{figure}
\centerline{
\psfig{file=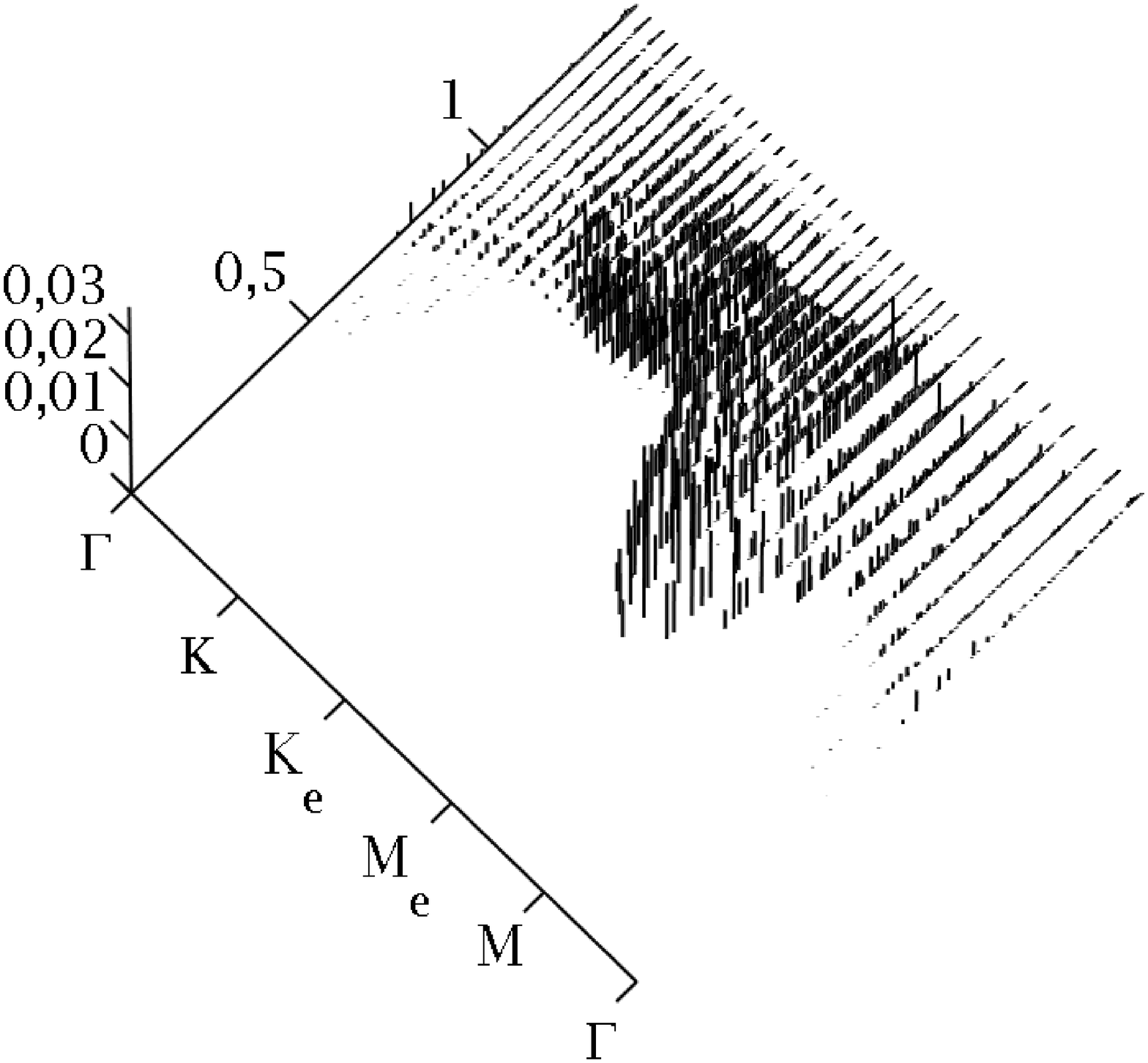,width=4cm,angle=0}
\psfig{file=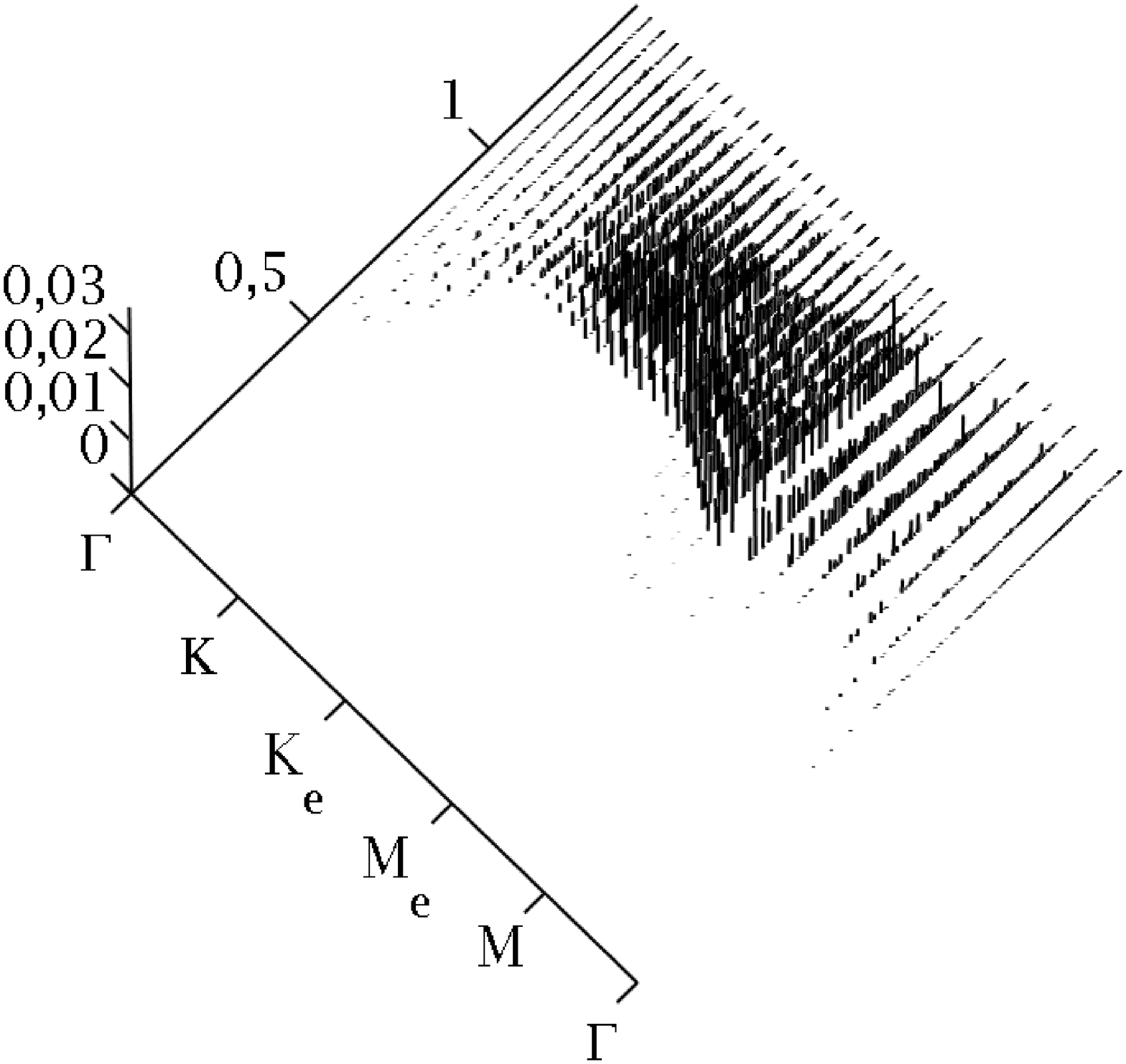,width=4cm,angle=0}}
\vspace{.5cm}
\centerline{
\psfig{file=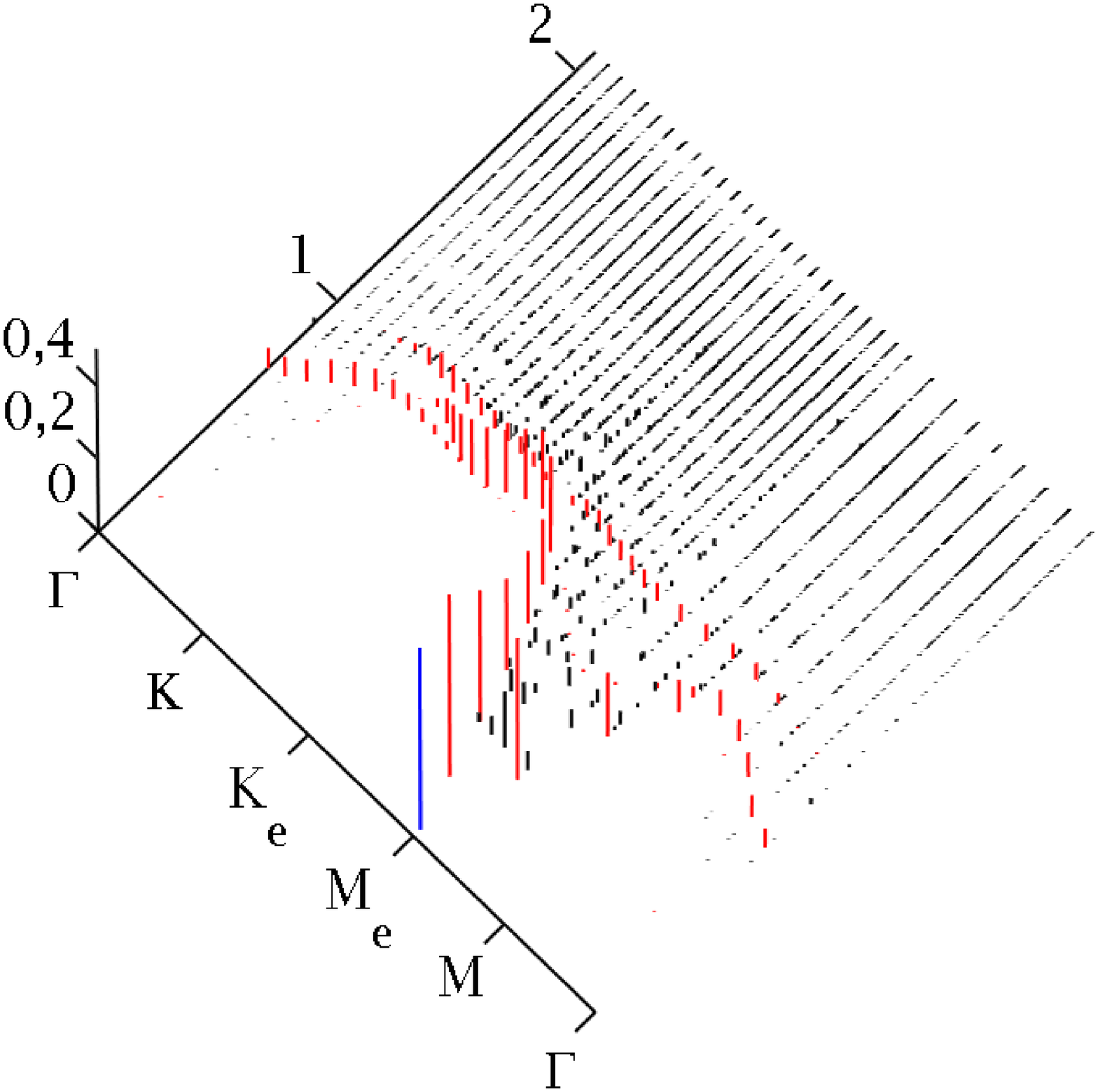,width=4cm,angle=0}
\psfig{file=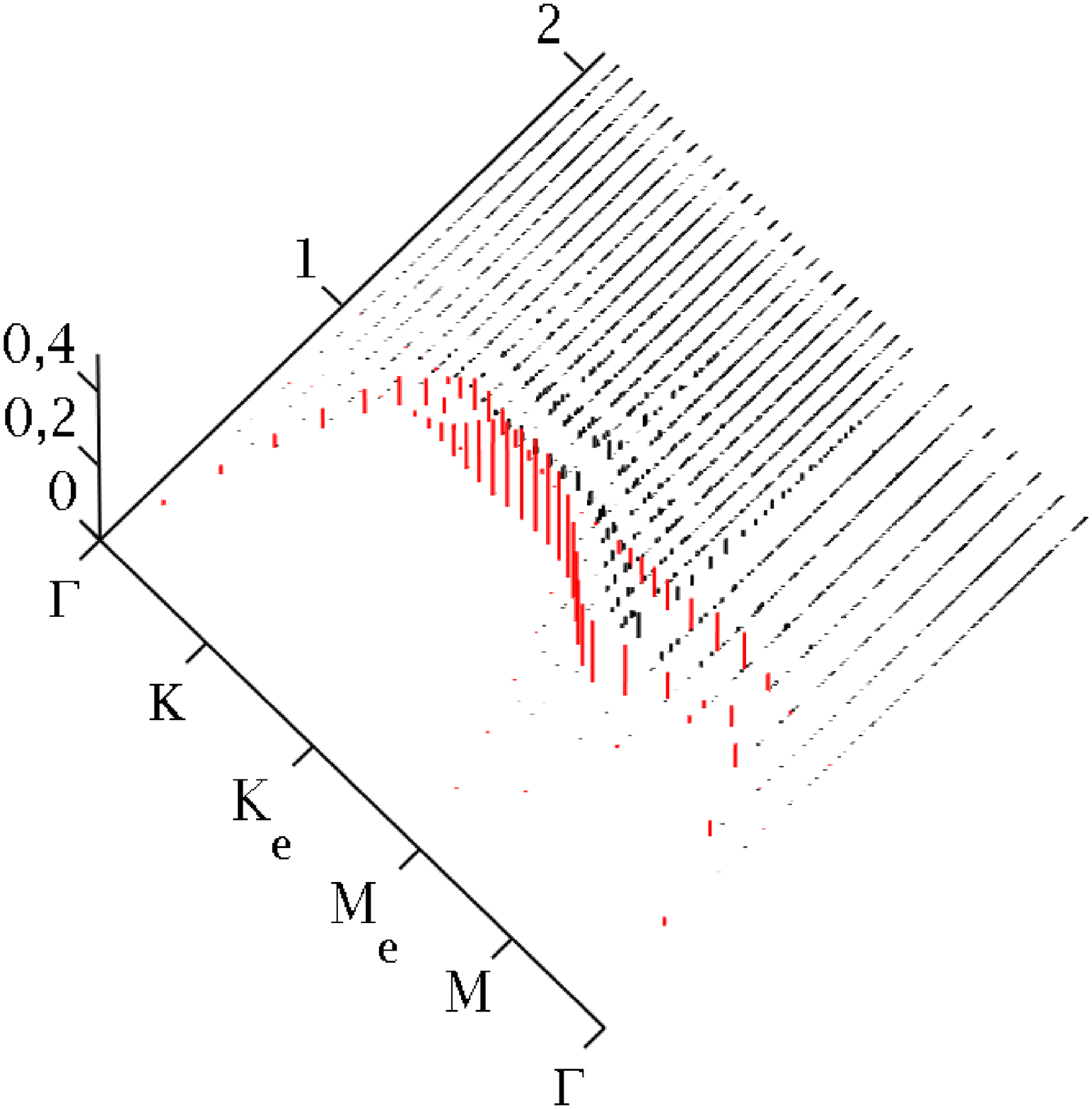,width=4cm,angle=0}
}
\caption{(Color online) Dynamical structure factors $S^{xx}(\mathbf{q},\omega)$ (left columns) and  $S^{zz}(\mathbf{q},\omega)$ (right columns) for the $Q=0$ \textit{Ansatz} in the spin-liquid phase, $S=0.1$ (top) or with long-range order, $S=0.2$ (bottom). The system size is $N=576$. $\theta$ is taken to be 0.25 to emphasize the anisotropy of  $S^{xx}(\mathbf{q},\omega)$  and  $S^{zz}(\mathbf{q},\omega)$. In the N\'eel phase, the largest  peak (in blue) is the quasi-elastic peak, next in intensity are the magnon branches (in red) (intensities are cut to see the weaker continuum).}
\label{dyn1}
\end{figure}

We now discuss the general features of the dynamical spin structure factor and show the results for the $Q=0$ \textit{Ansatz}  in Fig.~\ref{dyn1}.

In the spin liquid phase ($S<S_c(\theta)$),  all spinons are gapped and the two-spinon excitations form a gapped continuum, the bottom edge of which is given by the minimum of $ \omega_{\mathbf{Q+q} l} + \omega_{-\mathbf{q} n} $ over all $\mathbf{q}$ and  all spinon bands $(l, n)$.  $S^{xx}(\mathbf{Q},\omega)$ and $S^{zz}(\mathbf{Q},\omega)$ are given in Figs.~\ref{dyn1}~(top). In these figures, we have taken  $D$ large enough to see the anisotropy of the response (and a small $S=0.1$ to be in the spin liquid phase). When $S$ increases the lower edge of the continuum shifts to zero and its intensity increases continuously.

For $S>S_c(\theta)$ (Figs.~\ref{dyn1} (bottom), $S=0.2$, $\theta=0.25$), due to the soft spinon mode $\tilde{\phi}_{\mathbf{q}_0 l_0 \sigma}$ (See Fig.~\ref{fig:Espinons} (right))  the system enters a Bose-condensed phase: it has long-range correlations and  low-energy excitations varying as $1/N$, at $\mathbf{M}_e$. This spinon mode has a singular contribution $\sim \sqrt{N}$ [see  Eq.~(\ref{BEC})]. As a consequence, the intensities have different finite-size scaling, depending on whether the pair of excited spinons contains the soft spinon or not. We can therefore identify three contributions:

\begin{itemize}

\item \textit{Elastic peak}. This is the peak with the largest
  intensity at $\mathbf{Q}=\mathbf{M}_e$ [shown in blue (online) in
  Fig.~\ref{dyn1} (bottom left) and cut in intensity to show the other
  excitations].  It comes from a pair of (identical) soft spinons
  (with wave vector $\mathbf{q}_0 = 0$) in Eq.~(\ref{2spinons}), and
  so has energy ${\cal O}(1/N)$ and intensity proportional to
  $|U_{\mathbf{q}_0il_0}V_{\mathbf{q}_0il_0}/\sqrt{N}|^2 \sim N$. By
  integrating over all frequencies, only this (zero-frequency) peak
  contributes to the peak of the (equal-time) structure factor,
  $S^{xx}(\mathbf{M}_e)$ (Fig.~\ref{stat} (right)). We also note that
  the peak is absent in the $zz$ response (Fig.~\ref{dyn1} (bottom
  right)), which is expected because the correlations are long-ranged
  in the plane only.

\item \textit{Magnon branches}. There are three magnon branches (shown
  in red (online) in Figs.~\ref{dyn1} (bottom)), two are ``optical''
  modes, the third one, gapless, is the Goldstone mode of the U(1)
  symmetry. The (almost) flat mode is the \textit{weathervane mode}
  which is always gapped in the Schwinger boson approach, contrary to
  spin-wave theory,\cite{Harris} and irrespective of
  $S$.\cite{Sachdev} The magnon here consists of a pair of the soft
  spinon and a spinon of wave-vector $\mathbf{Q}$ [see
    Eq.~(\ref{2spinons})] so that the magnon dispersion \textit{is}
  the spinon dispersion $\omega_{\mathbf{Q}\mu}$,\cite{order} and the
  intensities are of order
  $|U_{\mathbf{q}in}V_{\mathbf{q}_0il_0}/\sqrt{N}|^2 \sim 1$.

\item \textit{Continuum}. As for the spin-liquid phase, there is a continuum of excitations obtained from contributions in (\ref{2spinons}) with two spinons both different from the soft mode. Each peak has intensity $|U_{\mathbf{q}in}V_{\mathbf{k}il}/\sqrt{N}|^2 \sim {\cal O}(1/N)$ and the sum of them gives a continuum with finite intensity in the thermodynamic limit, which is absent in lowest-order spin-wave theory.
\end{itemize}
All these excitations contribute to the
sum-rule, $N \langle (S_i^x)^2\rangle = \sum_q \int (d\omega/2\pi)
S^{xx}(q,\omega)$, given the different density of states.
Note that, as expected, the \DM interaction introduces an anisotropy between the in-plane (left column of Fig.~\ref{dyn1}) and the out-of-plane (right column of Fig.~\ref{dyn1}) dynamical factors. This anisotropy is visible in the spin-liquid as well as in the N\'eel ordered phase. The effect is more spectacular in the latter where the \DM interaction strongly suppresses the low-energy response in the $zz$ channel around the $\mathbf{M}_e$ point.

\section{Conclusion}

We have obtained the Schwinger-boson mean-field phase-diagram for
different values of $S$. The large $S$ limit is in agreement with the
semi-classical order by disorder mechanism which selects the $\sqrt{3}
\times \sqrt{3}$ state.\cite{Sachdev} We find that this phase remains
stable at small anisotropy in a region which becomes broader and
broader when the spin decreases (and hence quantum mechanical effects
increase). It is therefore possible in principle to observe both
ordered phases experimentally and first-order transitions between
them. However, given the small critical strength, the $Q=0$ phase is
more likely to occur in a real compound with large enough $S$ and the
kagome potassium jarosite ($S=5/2$) offers such an
example.\cite{Wills,singleion}

We have identified a region of
the phase-diagram ($S \sim 0.2$, \textit{Ansatz}
$(\pi,0)$) which resembles qualitatively to the exact diagonalization
results for the $S=1/2$ system, where the \DM interaction induces a quantum
phase transition between a topological spin liquid and the $Q=0$
N\'eel ordered phase. This suggests to consider smaller values of the
``spin'' parameter $S$ as possibly relevant within the Schwinger-boson
theory, given that the treatment on average of the constraint leads to enhance $\langle \mathbf{S}_i^2 \rangle$ compared to
$S(S+1)$.  Within this framework, we could calculate observables such
as (i) the cross-section of diffuse neutron scattering, with
the evolution from broad diffuse scattering to Bragg peaks across the
quantum phase transition (Figs.~\ref{stat} and \ref{powder}) (ii) the neutron inelastic
cross-section which, in addition to Bragg peaks and spin waves, shows
a broad continuum in both disordered and ordered phases
(Fig.~\ref{dyn1}), absent in the lowest-order of spin-wave theory.

\section*{Acknowledgments}
We thank G. Misguich and B. Dou\c cot for numerous enlightening comments on the Schwinger boson theory.


\begin{thebibliography}{99}
\bibitem{Gregoire} At finite $\cal N$ and for sufficiently small boson density,
the problem is strictly equivalent to gapped spinons (the Schwinger
bosons) interacting with a fluctuating bond
field. It has been understood that the low-energy physics
of such a system will be that of a lattice gauge theory coupled to
charged matter fields.
But the nature (group) of the gauge field crucially depends on the
lattice geometry, and in the present case where the lattice is
not bipartite, it should generically be of $\mathbb Z_2$ type, and has
two types of phases.\cite{ReadSachdev91,m08}
If the effective $\mathbb Z_2$ gauge theory is in a deconfined phase
(which should be the case
for large enough $\cal N$),  the ground-state is qualitatively close to
the mean-field one, with gapped and unconfined spinons.
If the phase is instead confined, the gauge fluctuations are strong and
cannot be neglected, the mean-field picture is not valid any
more.
\bibitem{simplicity} Fore the sake of simplicity we develop only the
formulas corresponding to a unique soft mode.
This is the case of the N\'eel order in presence of a \DM interaction: due to the $U(1)$ symmetry of the original Hamiltonian, the Goldstone mode is unique.
The general case is straightforward but more cumbersome in writing.
\bibitem{normalization} $m_{AF}^2$ is delicate to normalize because of
  the fluctuations of the spin length on each site. Fluctuations also
  increase with the strength of the \DM coupling, making the normalization partly arbitrary.
\bibitem{order} If $\mathbf{q}_0 \neq 0$, the spinon dispersion is folded in the magnetic Brillouin zone.
\bibitem{singleion} For $S>1/2$, we note that single-ion anisotropies
are also present and may affect the form of magnetic order.
\end{thebibliography}
\end{document}